\documentclass{ifacconf}

\usepackage{natbib}        

\usepackage{amsmath} 
\usepackage{amssymb}  
\usepackage{bbm}
\usepackage{color}
\usepackage{graphics}
\usepackage{graphicx}
\usepackage{epsfig}
\usepackage{epstopdf}
\usepackage{amsfonts}
\usepackage{mathtools}
\usepackage{bm}
\usepackage{fancyhdr}
\usepackage{framed}
\usepackage{float}
\usepackage{cite}
\usepackage{verbatim}
\usepackage{tikz}
\usetikzlibrary{calc,positioning,shapes,shadows,arrows,fit}

\newtheorem{problem}{Problem}

\newtheorem{lemma}{Lemma}
\newtheorem{proposition}{Proposition}
\newtheorem{remark}{Remark}
\newtheorem{definition}{Definition}

\newtheorem{assumption}{Assumption}

\begin{document}
\begin{frontmatter}

\title{Robust Distance-Based Formation Control of Multiple Rigid Bodies with Orientation Alignment\thanksref{footnoteinfo}} 

\thanks[footnoteinfo]{This work was supported by the H2020 ERC Starting Grand BUCOPHSYS, the Swedish Research Council (VR), the Knut och Alice Wallenberg Foundation and the European Union's Horizon 2020 Research and Innovation Programme under the Grant Agreement No. 644128 (AEROWORKS).}

\author[kth]{Alexandros Nikou} 
\author[kth]{Christos K. Verginis} 
\author[kth]{Dimos V. Dimarogonas}
\address[kth]{ACCESS Linnaeus Center, School of Electrical Engineering and KTH Center for Autonomous Systems, KTH Royal Institute of Technology, SE-100 44, Stockholm, Sweden. \\ E-mail: \{anikou, cverginis, dimos\}@kth.se}

\begin{abstract}                
This paper addresses the problem of distance- and orientation-based formation control of a class of second-order nonlinear multi-agent systems in $3$D space, under static and undirected communication topologies. More specifically, we design a decentralized model-free control protocol in the sense that each agent uses only local information from its neighbors to calculate its own control signal, without incorporating any knowledge of the model nonlinearities and exogenous disturbances. Moreover, the transient and steady state response is solely determined by certain designer-specified performance functions and is fully decoupled by the agents' dynamic model, the control gain selection, the underlying graph topology as well as the initial conditions. Additionally, by introducing certain inter-agent distance constraints, we guarantee collision avoidance and connectivity maintenance between neighboring agents. Finally, simulation results verify the  performance of the proposed controllers.
\end{abstract}


\end{frontmatter}

\section{Introduction}

During the last decades, decentralized control of networked multi-agent systems has gained a significant amount of attention due to the great variety of its applications, including  multi-robot systems, transportation, multi-point surveillance and biological systems. The main focus of multi-agent systems is the design of distributed control protocols in order to achieve global tasks, such as consensus \citep{ren_beard_consensus, olfati_murray_concensus, jadbabaie_morse_coordination, tanner_flocking}, and at the same time fulfill certain properties, e.g., network connectivity \citep{egerstedt_formation, zavlanos_2008_distributed}. 

A particular multi-agent problem that has been considered in the literature is the formation control problem, where the agents represent robots that aim to form a prescribed geometrical shape, specified by a certain set of desired relative configurations between the agents. The main categories of formation control that have been studied in the related literature are (\citep{oh_park_ahn_2015}) position-based control, displacement-based control, distance-based control and orientation-based control. Distance- and orientation-based control constitute the topics in this work.

In distance-based formation control, inter-agent distances are actively controlled to achieve a desired formation, dictated by desired inter-agent distances. Each agent is assumed to be able to sense the relative positions of its neighboring agents, without the need of orientation alignment of the local coordinate systems. When orientation alignment is considered as a control design goal, the problem is known as orientation-based (or bearing-based) formation control. The desired formation is then defined by relative inter-agent orientations. The orientation-based control steers the agents to configurations that achieve desired relative orientation angles. 
In this work, we aim to design a decentralized control protocol such that both distance- and orientation-based formation is achieved.

The literature in distance-based formation control is rich, and is traditionally categorized in single or double integrator agent dynamics and directed or undirected communication topologies (see e.g. \citep{olfati_murray_2002, smith_broucke_francis_2006, hendrickx_anderson_delvenne_blondel_2007, anderson-yu-dasgupta-morse_2007, anderson_yu_fidan_hendrickx_2008, dimos_kalle_2008, cao_anderson_morse_yu_2008, yu_anderson_dagsputa_fidan_2009, krick_broucke_francis_2009, dorfler_francis_2010, oh_ahn_2011e, cao_morse_yu_anderson_dagsputa_2011, summers_yu_dagsputa_anderson_2011_tac, park_oh_ahn_2012_gradient, belabbas2012robustness, oh_ahn_2014a})

Orientation-based formation control has been addressed in \citep{basiri_2010_angle_formation, eren_2012_bearing_formation, oh_ahn_2014_angle_based_formation, zhao2016bearing}, whereas the authors in \citep{oh_ahn_2014_angle_based_formation, bishop_2015_distributed, fathian2016globally} have considered the combination of distance- and orientation-based formation.

In most of the aforementioned works in formation control, the two-dimensional case with simple dynamics and point-mass agents has been dominantly considered. In real applications, however, the engineering systems have nonlinear second order dynamics and are usually subject to exogenous disturbances and modeling errors. Another important issue concerns the connectivity maintenance, the collision avoidance between the neighboring agents and the transient and steady state response of the closed loop system, which have not been taken into account in the majority of related woks. Thus, taking all the above into consideration, the design of robust distributed control schemes for the multi-agent formation control problem becomes a challenging task. 

Motivated by this, we aim to address here the distance-based formation control problem with orientation alignment for a team of rigid bodies operating in 3D space, with unknown second-order nonlinear dynamics and external disturbances. We propose a purely decentralized control protocol that guarantees distance formation, orientation alignment as well as collision avoidance and connectivity maintenance between neighboring agents and in parallel ensures the satisfaction of prescribed transient and steady state performance. The prescribed performance control framework has been incorporated in multi-agent systems in \citep{karayiannidis_consensus_ppc_2012, babis_2014_formation}, where first order dynamics have been considered. Furthermore, the first one only addresses the consensus problem, whereas the latter solves the position based formation control problem, instead of the distance- and orientation-based problem treated here.

The remainder of the paper is structured as follows. In Section \ref{sec:preliminaries} notation and preliminary background is given. Section \ref{sec:prob_formulation} provides the system dynamics and the formal problem statement. Section \ref{sec:solution} discusses the technical details of the solution and Section \ref{sec:simulation_results} is devoted to a simulation example. Finally, the conclusion and future work directions are discussed in Section \ref{sec:conclusions}.

\section{Notation and Preliminaries} \label{sec:preliminaries}

\subsection{Notation} 

The set of positive integers is denoted as $\mathbb{N}$. The real $n$-coordinate space, with $n\in\mathbb{N}$, is denoted as $\mathbb{R}^n$;
$\mathbb{R}^n_{\geq 0}$ and $\mathbb{R}^n_{> 0}$ are the sets of real $n$-vectors with all elements nonnegative and positive, respectively. Given a set $S$, we denote as $\lvert S\lvert$ its cardinality. The notation $\|x\|$ is used for the Euclidean norm of a vector $x \in \mathbb{R}^n$. Given a symmetric matrix $A, \lambda_{\text{min}}(A) = \min \{|\lambda| : \lambda \in \sigma(A) \}$ denotes the minimum eigenvalue of $A$, respectively, where $\sigma(A)$ is the set of all the eigenvalues of $A$ and $rank(A)$ is its rank; $A \otimes B$ denotes the Kronecker product of matrices $A, B \in \mathbb{R}^{m \times n}$, as was introduced in \citep{horn_jonshon}. Define by $\mathbbm{1}_n \in \mathbb{R}^n, I_n \in \mathbb{R}^{n \times n}, 0_{m \times n} \in \mathbb{R}^{m \times n}$ the column vector with all entries $1$, the unit matrix and the $m \times n$ matrix with all entries zeros, respectively; $\mathcal{B}(c,r) = \{x \in \mathbb{R}^3: \|x-c\| \leq r\}$ is the $3$D sphere of radius $r\geq 0$ and center $c\in\mathbb{R}^{3}$. 
The vector connecting the origins of coordinate frames $\{A\}$ and $\{B$\} expressed in frame $\{C\}$ coordinates in $3$D space is denoted as $p^{\scriptscriptstyle C}_{{\scriptscriptstyle B/A}}\in{\mathbb{R}}^{3}$. Given $a\in\mathbb{R}^3$, $S(a)$ is the skew-symmetric matrix
defined according to $S(a)b = a\times b$. We further denote as $q_{\scriptscriptstyle B/A}\in\mathbb{T}^3$ the Euler angles representing the orientation of frame $\{B\}$ with respect to frame $\{A\}$, where $\mathbb{T}^3$ is the $3$D torus. The angular velocity of frame $\{B\}$ with respect to $\{A\}$, expressed in frame $\{C\}$ coordinates, is
denoted as $\omega^{\scriptscriptstyle C}_{\scriptscriptstyle B/A}\in \mathbb{R}^{3}$. We also use the notation $\mathbb{M} = \mathbb{R}^3\times \mathbb{T}^3$. For notational brevity, when a coordinate frame corresponds to an inertial frame of reference $\{0\}$, we will omit its explicit notation (e.g., $p_{\scriptscriptstyle B} = p^{\scriptscriptstyle 0}_{\scriptscriptstyle B/0}, \omega_{\scriptscriptstyle B} = \omega^{\scriptscriptstyle 0}_{\scriptscriptstyle B/0}$ etc.). All vector and matrix differentiations are derived with respect to an inertial frame $\{0\}$, unless otherwise stated.

\subsection{Prescribed Performance Control}
\label{subsec:ppc}
\begin{figure}
\centering
  \includegraphics[width=0.3\textwidth]{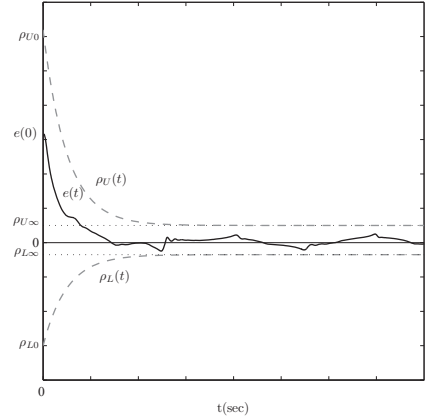}
\caption{Graphical illustration of the prescribed performance definition.}
\label{fig:ppc}       
\end{figure}

Prescribed Performance control, originally proposed in \citep{bechlioulis_tac_2008}, describes the behavior where a tracking error $e(t):\mathbb{R}_{\geq 0} \to \mathbb{R}$ evolves strictly within a predefined region that is bounded by certain functions of time, achieving prescribed transient and steady state performance.
The mathematical expression of prescribed performance is given by the following inequalities:
\begin{equation}
-\rho_L(t) < e(t) < \rho_U(t),\ \ \forall t\in\mathbb{R}_{\geq 0}, \notag
\end{equation} 
where $\rho_L(t),\rho_U(t)$ are smooth and bounded decaying functions of time, satisfying $\lim\limits_{t \to \infty}\rho_L(t) > 0$ and $\lim\limits_{t \to \infty}\rho_U(t) > 0$, called performance functions (see Fig. \ref{fig:ppc}). Specifically, for the exponential performance functions $\rho_i(t) = (\rho_{i 0}-\rho_{i \infty})e^{-l_it}+\rho_{i \infty}$, with $\rho_{i 0},\rho_{i \infty}, l_i\in\mathbb{R}_{>0}, i\in\{U,L\}$, appropriately chosen constants, $\rho_{L 0}=\rho_L(0),\rho_{U 0}=\rho_U(0)$ are selected such that $\rho_{U 0} > e(0) > \rho_{L 0}$ and the constants $\rho_{L \infty}=\lim\limits_{t \to \infty}\rho_L(t)<\rho_{L 0},\rho_{U \infty}=\lim\limits_{t \to \infty}\rho_U(t)<\rho_{U 0}$ represent the maximum allowable size of the tracking error $e(t)$ at steady state, which may be set arbitrarily small to a value reflecting the resolution of the measurement device, thus achieving practical convergence of $e(t)$ to zero. Moreover, the decreasing rate of $\rho_L(t),\rho_U(t)$, which is affected by the constants $l_L, l_U$ in this case, introduces a lower bound on the required speed of convergence of $e(t)$. Therefore, the appropriate selection of the performance functions $\rho_L(t),\rho_U(t)$ imposes performance characteristics on the tracking error $e(t)$.

\subsection{Dynamical Systems} \label{subsec:dynamical systems}
Consider the initial value problem:
\begin{equation}
\dot{\psi} = H(t,\psi), \psi(0)=\psi^0\in\Omega_{\psi}, \label{eq:initial_value_problem}
\end{equation}
with $H:\mathbb{R}_{\geq 0}\times\Omega_{\psi} \to \mathbb{R}^n$, where $\Omega_{\psi}\subseteq\mathbb{R}^n$ is a non-empty open set.
\begin{definition} (\citep{sontag_2013_mathematical})
A solution $\psi(t)$ of the initial value problem \eqref{eq:initial_value_problem} is maximal if it has no proper right extension that is also a solution of \eqref{eq:initial_value_problem}. 
\end{definition}
\begin{thm} (\citep{sontag_2013_mathematical}) \label{thm:dynamical systems}
Consider the initial value problem \eqref{eq:initial_value_problem}. Assume that $H(t,\psi)$ is: a) locally Lipschitz in $\psi$ for almost all $t\in\mathbb{R}_{\geq 0}$, b) piecewise continuous in $t$ for each fixed $\psi\in\Omega_{\psi}$ and c) locally integrable in $t$ for each fixed $\psi\in\Omega_{\psi}$. Then, there exists a maximal solution $\psi(t)$ of \eqref{eq:initial_value_problem} on the time interval $[0,\tau_{\max})$, with $\tau_{\max}\in\mathbb{R}_{> 0}$ such that $\psi(t)\in\Omega_{\psi},\forall t\in[0,\tau_{\max})$.
\end{thm}
\begin{proposition} (\citep{sontag_2013_mathematical}) \label{prop:dynamical systems}
Assume that the hypotheses of Theorem \ref{thm:dynamical systems} hold. For a maximal solution $\psi(t)$ on the time interval $[0,\tau_{\max})$ with $\tau_{\max}<\infty$ and for any compact set $\Omega'_{\psi}\subseteq\Omega_{\psi}$, there exists a time instant $t'\in[0,\tau_{\max})$ such that $\psi(t')\notin\Omega'_{\psi}$.
\end{proposition}

\subsection{Graph Theory}
An \textit{undirected graph} $\mathcal{G}$ is a pair $(\mathcal{V}, \mathcal{E})$, where $\mathcal{V}$ is a finite set of nodes, representing a team of agents, and $\mathcal{E} \subseteq \{ \{i,j\} : i,j \in \mathcal{V}, i \neq j\}$, with $M = |\mathcal{E}|$, is the set of edges that model the communication capability between neighboring agents. For each agent, its neighbors' set $\mathcal{N}_i$ is defined as $\mathcal{N}_i = \{j_1, \ldots, j_{N_i} \} = \{ j \in \mathcal{V} \text{ s.t. } \{i,j\} \in \mathcal{E}\}$, where $N_i = |\mathcal N_i|$.

If there is an edge $\{i, j\} \in \mathcal{E}$, then $i, j$ are called \textit{adjacent}. A \textit{path} of length $r$ from vertex $i$ to vertex $j$ is a sequence of $r+1$ distinct vertices, starting with $i$ and ending with $j$, such that consecutive vertices are adjacent. For $i = j$, the path is called a \text{cycle}. If there is a path between any two vertices of the graph $\mathcal{G}$, then $\mathcal{G}$ is called \textit{connected}. A connected graph is called a \text{tree} if it contains no cycles.

The \textit{adjacency matrix} $A(\mathcal{G}) = [a_{ij}] \in\mathbb{R}^{N \times N}$ of graph $\mathcal{G}$ is defined by $a_{ij}=a_{ji} = 1$, if $\{i,j\} \in \mathcal{E}$, and $a_{ij} =0$ otherwise. The \textit{degree} $d(i)$ of vertex $i$ is defined as the number of its neighboring vertices, i.e. $d(i) = N_i, i \in \mathcal{V}$. Let also $\Delta(\mathcal{G}) = \text{diag} \{[ d(i)]_{i\in\mathcal{V}}\} \in \mathbb{R}^{N \times N}$ be the \textit{degree matrix} of the system. Consider an arbitrary orientation of $\mathcal{G}$, which assigns to each edge $\{i, j\} \in \mathcal{E}$ precisely one of the ordered pairs $(i, j)$ or $(j, i)$. When selecting the pair $(i, j)$, we say that $i$ is the tail and $j$ is the head of the edge $\{i, j\}$. By considering a numbering $k \in \mathcal{M} = \{1, . . . , M\}$ of the graph's edge set, we define the $N \times M$ \textit{incidence matrix} $D(G)$ as it was given in \citep{mesbahi_2010_graph_theory}. The \textit{Laplacian matrix} $L(\mathcal{G}) \in \mathbb{R}^{N \times N}$ of the graph $\cal{G}$ is defined as $L(\mathcal{G}) = \Delta(\mathcal{G}) - A(\mathcal{G}) = D(\mathcal{G}) D(\mathcal{G})^{\tau}$.
\begin{lemma} \label{lemma:tree} \citep[Section III]{dimos_kalle_2008}
Assume that the graph $\mathcal{G}$ is a tree. Then, $D^\tau(\mathcal{G})D(\mathcal{G})$ is positive definite.
\end{lemma}
\section{Problem Formulation} \label{sec:prob_formulation}
\subsection{System Model}
\begin{figure}[ht!]
	\centering
	\begin{tikzpicture}[scale = 0.5]	
	\draw [color=black,thick,->,>=stealth'](-9, -5) to (-7, -5);
	\draw [color=black,thick,->,>=stealth'](-9, -5) to (-9, -3);
	\draw [color=black,thick,->,>=stealth'](-9, -5) to (-10, -6.5);
    \node at (-9.8, -5.0) {$\{0\}$};
    
    \draw [color = blue, fill = blue!20] (-4.5,0) circle (2.5cm);
    \node at (-5.7, 0.0) {$\{i\}$};
    \draw[green,thick,dashed] (-4.5,0) circle (5.0cm);
    \draw [color=black,thick,->,>=stealth'](-9, -5) to (-4.5, -0.1);
    \node at (-7.7, -3.0) {$p_i$};
    \draw [color=green,thick,dashed,->,>=stealth'](-4.5, 0.0) to (-8.93, 2.43);
    \node at (-7.3, 2.15) {$s_i$};
    \draw [color=black,thick,dashed,->,>=stealth'](-4.5, 0.0) to (-2.0, 0.0);
    \node at (-3.3, 0.3) {$r_i$};
    \node at (-4.5, 0.0) {$\bullet$};
    
    \draw [color = red, fill = red!20] (3.2, 0) circle (1.5cm);
    \node at (2.5, 0.3) {$\{j\}$};
    \draw[orange,thick,dashed,] (3.2, 0) circle (4.1cm);
    \draw [color=black,thick,->,>=stealth'](-9, -5) to (3.2, -0.1);
    \node at (-5.0, -3.9) {$p_j$};
    \draw [color=orange,thick,dashed,->,>=stealth'](3.2, 0.0) to (3.2, -4.0);
    \node at (3.8, -2.7) {$s_j$};
    \draw [color=black,thick,dashed,->,>=stealth'](3.2, 0.0) to (4.7, 0.0);
    \node at (4.1, 0.3) {$r_j$};
    \node at (3.2, 0.0) {$\bullet$};
	\end{tikzpicture}
	\caption{Illustration of two agents $i, j \in \mathcal{V}$ in the workspace; $\{0\}$ is the inertial frame, $\{i\}, \{j\}$ are the frames attached to the agents' center of mass, $p_i, p_j \in \mathbb{R}^3$ are the positions of the center of mass with respect to $\{0\}$, $r_i, r_j$ are the radii of the agents and $s_i > s_j$ are their sensing ranges.}
	\label{fig:agents_geometry}
\end{figure}
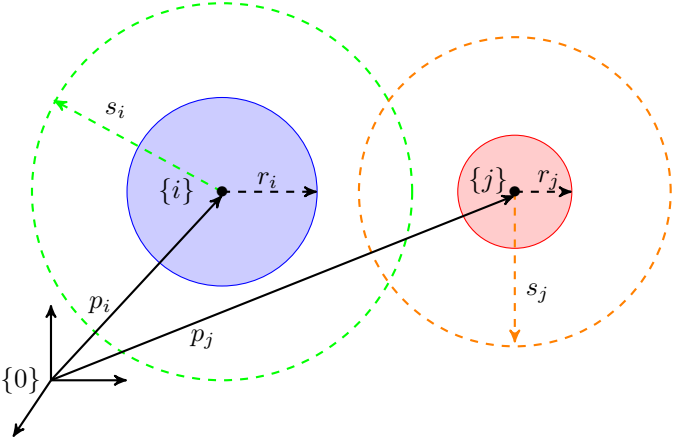

Consider a set of $N$ rigid bodies, with $\mathcal{V} = \{ 1,2, \ldots, N\}$, $N  \geq 2$, operating in a workspace $W\subseteq \mathbb{R}^3$, with coordinate frames $\{i\}, i\in\mathcal{V}$, attached to their centers of mass. We consider that each agent occupies a sphere $\mathcal{B}_{r_i}(p_i(t))$, where $p_i:\mathbb{R}_{\geq 0} \to \mathbb{R}^3$ is the position of the agent's center of mass and $r_i$ is the agent's radius (see Fig. \ref{fig:agents_geometry}). We also denote as $q_i:\mathbb{R}_{\geq 0} \to \mathbb{T}^3, i\in\mathcal{V}$, the Euler angles representing the agents' orientation with respect to an inertial frame $\{0\}$, with $q_i = [\phi_i,\theta_i,\psi_i]^\tau$. By defining $x_i:\mathbb{R}_{\geq 0} \to \mathbb{M}, v_i : \mathbb{R}_{\geq 0} \to \mathbb{R}^6,$ with $x_i = [p^\tau_i,q^\tau_i]^\tau, v_i =[\dot{p}^\tau_i,\omega^\tau_i]^\tau$, we model each agent's motion with the $2$nd order dynamics:
\begin{subequations}\label{eq:system} 
\begin{align} 
& \dot{x}_i(t) = J_i(x_i)v_i(t) , \label{eq:system_1} \\ 
& M_i(x_i) \dot{v}_i(t) + C_i(x_i,\dot{x}_i) v_i(t)+g_i(x_i) \notag \\
&\hspace{43mm}+ w_i(x_i,\dot{x}_i,t)= u_i,  \label{eq:system_2} 
\end{align}
\end{subequations}
where $J_i:\mathbb{M} \to \mathbb{R}^{6\times6}$ is a Jacobian matrix that maps the Euler angle rates to $v_i$, given by 
\begin{align} 
J_i(x_i) 
&=
\begin{bmatrix}
I_3 & 0_{3 \times 3} \\
0_{3 \times 3} & J_{ q }(x_i) \\
\end{bmatrix} \notag, \\
J_q(x_i) 
&= 
\begin{bmatrix}
1 & \sin(\phi_i) \tan(\theta_i)  & \cos(\phi_i) \tan(\theta_i) \\
  0 & \cos(\phi_i) & -\sin(\phi_i) \\
 0  & \displaystyle \frac{\sin(\phi_i)}{\cos(\theta_i)} & \displaystyle \frac{\cos(\phi_i)}{\cos(\theta_i)}
\end{bmatrix}, \notag
\end{align}
for which we make the following assumption:
\begin{assumption} \label{as:J}
The angle $\theta_i$ satisfies the inequality $-\frac{\pi}{2} < \theta_i(t) < \frac{\pi}{2} ,\forall i\in\mathcal{V},t\in\mathbb{R}_{\geq 0}$.
\end{assumption}
The aforementioned assumption guarantees that $J_i$ is always well-defined and invertible, since $\det(J_i) = \tfrac{1}{\cos\theta_i}$.
Furthermore, $M_i:\mathbb{M} \to \mathbb{R}^{6\times6}$ is the positive definite inertia matrix, $C_i:\mathbb{M}\times\mathbb{R}^6 \to \mathbb{R}^{6\times6}$ is the Coriolis matrix, $g_i:\mathbb{M} \to \mathbb{R}^6$ is the gravity vector, and $w_i:\mathbb{M}\times\mathbb{R}^{6}\times\mathbb{R}_{\geq 0} \to \mathbb{R}^6$ is a bounded vector representing model uncertainties and external disturbances. We consider that the aforementioned vector fields are unknown and continuous. 
Finally, $u_i\in\mathbb{R}^6$ is the control input vector representing the $6$D generalized force acting on the agent. 

The dynamics \eqref{eq:system} can be written in vector form as:
\begin{subequations}\label{eq:system_MAS} 
\begin{align} 
& \dot{x}(t) = J(x)v(t) , \label{eq:system_1_MAS} \\ 
& \bar{M}(x) \dot{v}(t) + \bar{C}(x,\dot{x}) v(t)+\bar{g}(x) + \bar{w}(x,\dot{x},t)= u	,  \label{eq:system_2_MAS} 
\end{align}
\end{subequations}
where $x = [x_1^\tau,\dots,x_N^\tau]^\tau : \mathbb{R}_{\geq 0} \to \mathbb{M}^N, v = [v_1^\tau, \dots, v_N^\tau]^\tau : \mathbb{R}_{\geq 0} \to \mathbb{R}^{6N} , u = [u_1^\tau, \dots, u_N^\tau]^\tau \in \mathbb{R}^{6N}$, and
\begin{align*}
J &= \text{diag}\{[J_i]_{i\in\mathcal{V}} \} \in\mathbb{R}^{6N\times 6N}, \\
\bar{M} &= \text{diag}\{[M_i]_{i\in\mathcal{V}} \} \in\mathbb{R}^{6N\times 6N}, \\
\bar{C} &= \text{diag}\{[C_i]_{i\in\mathcal{V}} \} \in\mathbb{R}^{6N\times 6N}, \\
\bar{g} &= [g_1^\tau,\dots, g^\tau_N]^\tau \in\mathbb{R}^{6N},  \\ 
\bar{w} &= [w_1^\tau,\dots, w^\tau_N]^\tau \in\mathbb{R}^{6N}. 
\end{align*}

It is also further assumed that each agent can measure its own $p_i,q_i, \dot{p}_i, v_i, i\in\mathcal{V}$, and has a limited sensing range of $s_i > \max\{r_i+r_j: i,j \in \mathcal{V}\}$. Therefore, by defining the neighboring set $\mathcal{N}_i(t) = \{j\in\mathcal{V} : p_j(t)\in\mathcal{B}_{s_i}(p_i(t))\}$, agent $i$ also knows at each time instant $t$ all $p^i_{j/i}(t), q_{j/i}(t)$ and, since it knows its own $p_i(t),q_i(t)$, it can compute all $p_{j}(t), q_{j}(t), \forall j\in \mathcal{N}_i(t),t\in\mathbb{R}_{\geq 0}$.

The topology of the multi-agent network is modeled through the graph $\mathcal{G} = (\mathcal{V},\mathcal{E})$, with $\mathcal{V}=\{1,\dots,N\}$ and $\mathcal{E}=\{\{i,j\}\in\mathcal{V}\times\mathcal{V} \text{ s.t. } j\in\mathcal{N}_i(0) \text{ and } i\in\mathcal{N}_j(0)\}$. The latter implies that at $t=0$ the graph is undirected, i.e., 
\begin{equation}
\lVert p_{\ell_k}(0)-p_{m_k}(0)\rVert < d_{k,\text{con}}, \forall \{\ell_k,m_k\}\in\mathcal{E},
\end{equation}
with $d_{k,\text{con}} = \min\{s_{\ell_k},s_{m_k}\},\ell_k,m_k\in\mathcal{V},\forall k\in\mathcal{M}$. We also consider that $\mathcal{G}$ is static in the sense that no edges are added to the graph. We do not exclude, however, edge removal through connectivity loss between initially neighboring agents, which we guarantee to avoid, as presented in the sequel. It is also assumed that at $t=0$ the neighboring agents are at a collision-free configuration, i.e., $d_{k,\text{col}} < \lVert p_{\ell_k}(0)-p_{m_k}(0)\rVert, \forall \{\ell_k,m_k\}\in\mathcal{E}$, with $d_{k,\text{col}} =r_{\ell_k}+r_{m_k}$. Hence, we conclude that 
\begin{equation}
d_{k,\text{col}} < \lVert p_{\ell_k}(0)-p_{m_k}(0)\rVert < d_{k,\text{con}}, \forall \{\ell_k,m_k\}\in\mathcal{E}. \label{eq:at t=0}
\end{equation}

Moreover, given the desired formation constants $d_{k,\text{des}}$, $q_{k,\text{des}}$ for the edge $k\in\mathcal{M}$, the formation configuration is called \textit{feasible} if the set ${\Phi} = \{x\in\mathbb{M}^N : \lVert p_{\ell_k} - p_{m_k} \rVert = d_{k,\text{des}}, q_{\ell_k} - q_{m_k} = q_{k,\text{des}}, \forall \{\ell_k,m_k\}\in\mathcal{E} \}$, with $\ell_k,m_k\in\mathcal{V},\forall k\in\mathcal{M}$, is nonempty.   
\subsection{Problem Statement}
Due to the fact that the agents are not dimensionless and their communication capabilities are limited, the control protocol, except from achieving a desired inter-agent formation, should also guarantee for all $t\in\mathbb{R}_{\geq 0}$ that (i) the neighboring agents avoid collision with each other and (iii) all the initial edges are maintained, i.e., connectivity maintenance. Therefore, all pairs $\{\ell_k,m_k\}\in\mathcal{V}\times\mathcal{V}$ of agents that initially form an edge must remain within distance greater than $d_{k,\text{col}}$ and less than $d_{k,\text{con}}$. We also make the following assumptions that are required on the graph topology: 
\begin{assumption} \label{assump:basic_assumption}
The communication graph $\mathcal{G}$ is initially a tree.
\end{assumption}
Formally, the robust formation control problem under the aforementioned constraints is formulated as follows:
\begin{problem} \label{problem}
	Given $N$ agents governed by the dynamics \eqref{eq:system}, under the Assumptions 1-2 and given the desired inter-agent distances and angles $d_{k,\text{des}}, q_{k,\text{des}}$, with $d_{k,\text{col}}<d_{k,\text{des}} < d_{k,\text{con}}$, $\forall \{\ell_k,m_k\}\in \mathcal{E}, \ell_k,m_k\in\mathcal{V}, \forall k\in\mathcal{M}$, design decentralized control laws $u_i \in\mathbb{R}^6,i\in\mathcal{V}$ such that $\forall \ \{\ell_k,m_k\}\in \mathcal{E}, k \in \mathcal{M}$, the following hold:
	\begin{enumerate}
	\item $\lim\limits_{t \to \infty} \|p_{\ell_k}(t)-p_{m_k}(t)\| = d_{k,\text{des}}$,
	\item $\lim\limits_{t \to \infty} \left[q_{m_k}(t) - q_{\ell_k}(t) - q_{k,\text{des}}\right] = 0_{3\times1}$, 
	\item $d_{k,\text{col}} < \|p_{\ell_k}(t)-p_{m_k}(t)\| < d_{k,\text{con}}, \forall \ t \in \mathbb{R}_{\geq 0}$.
	\end{enumerate}
\end{problem}
\section{Problem Solution} \label{sec:solution}
\subsection{Error Derivation} \label{subsec:error derivation}
Let $p = [p_1^\tau, \dots, p_N^\tau ]^\tau : \mathbb{R}_{\geq 0} \to \mathbb{R}^{3 N}, q = [q_1^\tau, \dots, q_N^\tau ]^\tau : \mathbb{R}_{\geq 0 } \to \mathbb{T}^{3N}$ be the stacked vectors of all the agent positions and Euler angles. We denote by $\tilde{p}, \tilde{q} :\mathbb{R}_{\geq 0 } \to \mathbb{R}^{3 M}$ the stack column vector of $p_{\ell_k,m_k}(t)=p_{\ell_k}(t)-p_{m_k}(t)$ and $q_{\ell_k,m_k}(t)=q_{\ell_k}(t)-q_{m_k}(t)$, respectively, $\forall\{\ell_k, m_k\} \in \mathcal{E}$, with the edges ordered as in the case of the incidence matrix $D(\mathcal{G})$. Thus, the following holds:
\begin{subequations} \label{eq:p q tilde}
\begin{align} 
\tilde{p}(t) &= 
\begin{bmatrix}
p_{\ell_1,m_1}(t) \\
\vdots \\
p_{\ell_M,m_M}(t) \\
\end{bmatrix}
= \begin{bmatrix} p_{\ell_1}(t)-p_{m_1}(t) \\ \vdots \\p_{\ell_M}(t)-p_{m_M}(t) \end{bmatrix} \notag \\
&= \left(D^\tau(\mathcal{G}) \otimes I_3 \right) p(t), \label{eq:p tilde} \\
\tilde{q}(t) &= \begin{bmatrix} q_{\ell_1}(t)-q_{m_1}(t) \\ \vdots \\q_{\ell_M}(t)-q_{m_M}(t)   \end{bmatrix} = \left(D^\tau(\mathcal{G}) \otimes I_3 \right) q(t). \label{eq:q tilde}
\end{align}
\end{subequations}
Next, let us introduce the errors $e^p_k:\mathbb{R}_{\geq 0} \to \mathbb{R}, e^q_k=[e^q_{k_1},e^q_{k_2},e^q_{k_3}]^\tau:\mathbb{R}_{\geq 0} \to \mathbb{T}^3$:
\begin{align}
 e^p_k(t) &= \left\|p_{\ell_k,m_k}(t) \right\|^2-d_{k,\text{des}}^2,  \notag\\
 e^q_k(t) &= q_{m_k}(t) - q_{\ell_k}(t) - q_{k,\text{des}}, \notag
\end{align}
for all distinct edges $\{\ell_k,m_k\} \in \mathcal{E}, k \in \mathcal{M}$, in the numbered order they appear in the edge set $\mathcal{E}$. 

By taking the time derivative of the aforementioned errors, the following is obtained:
\begin{subequations}
\begin{align}
&\hspace{-3mm}\dot{e}^p_k(t) = 2p^\tau_{\ell_k,m_k}(t) \dot{p}_{\ell_k,m_k}(t) \label{eq:e_p_k_deriv}, \\ 
&\hspace{-3mm}\dot{e}^q_k(t) = \dot{q}_{m_k}(t) - \dot{q}_{\ell_k}(t).  \label{eq:e_q_k_deriv} 
\end{align}
\end{subequations}
Also, by defining the vectors $e^p(t) = [e^p_1(t), \dots, e^p_M(t)]^{\tau}\in\mathbb{R}^M, e^q(t) = [(e^q_1(t))^\tau, \dots, (e^q_M(t))^\tau]^{\tau}\in\mathbb{T}^{3M}$ and employing \eqref{eq:p q tilde}, \eqref{eq:e_p_k_deriv} and \eqref{eq:e_q_k_deriv} can be written in vector form as:
\begin{subequations}  \label{eq:e_deriv}
\begin{align}
\dot{e}^p(t) 
&= 
\begin{bmatrix}
\dot{e}^p_1(t) \\
\vdots \\
\dot{e}^p_M(t)
\end{bmatrix} =
\begin{bmatrix}
2p^\tau_{\ell_1,m_1}(t)\dot{p}_{\ell_1,m_1}(t) \\
\vdots \\
2p^\tau_{\ell_M,m_M}(t)\dot{p}_{\ell_M,m_M}(t)
\end{bmatrix} \notag \\
&=
2 \begin{bmatrix}
p^\tau_{\ell_1,m_1}(t)  & \dots & 0_{1\times 3} \\
\vdots & \ddots & \vdots\\
0_{1\times 3} &  \dots  & p^\tau_{\ell_M,m_M}(t) \\
\end{bmatrix}
\begin{bmatrix}
\dot{p}_{\ell_1,m_1}(t) \\
\vdots \\
\dot{p}_{\ell_M,m_M}(t)
\end{bmatrix} \notag \\
&= 
\mathbb{F}_p(x) \left(D^{\tau}(\mathcal{G}) \otimes I_3 \right) \dot{p}, \label{eq:e_p_deriv} \\ \notag
\end{align}
\begin{align}
\dot{e}^q(t) =
\begin{bmatrix}
\dot{e}^q_1(t) \\
\vdots \\
\dot{e}^q_M(t)
\end{bmatrix} =
\begin{bmatrix}
\dot{q}_{\ell_1} - \dot{q}_{m_1} \\
\vdots \\
\dot{q}_{\ell_M} - \dot{q}_{m_M} 
\end{bmatrix} = \left(D^{\tau}(\mathcal{G}) \otimes I_3 \right)\dot{q}, \label{eq:e_q_deriv} 
\end{align}
\end{subequations}
where $\mathbb{F}_p:\mathbb{M}^N \to \mathbb{R}^{M\times 3M}$, with 
\begin{equation}
\mathbb{F}_p(x) = 2
\begin{bmatrix}
p^\tau_{\ell_1,m_1}(t)  & \dots & 0_{1\times 3} \\
\vdots & \ddots & \vdots\\
0_{1\times 3} &  \dots  & p^\tau_{\ell_M,m_M}(t) 
\end{bmatrix}. \notag 
\end{equation}
By introducing the stack error vector $e(t)=[(e^p(t))^\tau, \\ (e^q(t))^\tau]^\tau \in \mathbb{R}^{4M}$, \eqref{eq:e_deriv} can be written as:
\begin{align}
\dot{e}(t) = \bar{\mathbb{F}}_p(x) \bar{D}^\tau(\mathcal{G})
\begin{bmatrix}
\dot p \\
\dot q \\
\end{bmatrix}
, \label{eq:e_stack_deriv_a}
\end{align} 
where 
\begin{subequations} \label{eq:bar_F_bar_D}
\begin{align}
\bar{\mathbb{F}}_p(x) &= 
\begin{bmatrix}
\mathbb{F}_p(x) & 0_{M \times 3M} \\
0_{3M \times 3M} & I_{3M} \\
\end{bmatrix} \in \mathbb{R}^{4M \times 6M}, \label{eq:bar_F_p} \\
\bar{D}(\mathcal{G}) &= 
\begin{bmatrix}
D(\mathcal{G}) \otimes I_3 & 0_{3N\times 3M} \\  0_{3N\times 3M} & D(\mathcal{G})\otimes I_3 \label{eq:bar_D}
\end{bmatrix} \in\mathbb{R}^{6N\times 6M}.
\end{align}
\end{subequations}
Finally, we obtain from \eqref{eq:system_1_MAS}:
\begin{align}
\begin{bmatrix}
\dot p \\
\dot q \\
\end{bmatrix}
&=
\underbrace{\left[
\begin{array}{c c c|c c c}
I_3 & \dots & 0_{3\times 3} & 0_{3\times 3} & \dots & 0_{3\times 3} \\
\vdots & \ddots & \vdots & \vdots & \ddots & \vdots \\
0_{3\times 3} & \dots & I_3 & 0_{3\times 3} & \dots & 0_{3\times 3} \\
\hline
0_{3\times 3} & \dots & 0_{3\times 3} & J_{q}(x_1) & \dots & 0_{3\times 3} \\
\vdots & \ddots & \vdots & \vdots & \ddots & \vdots \\
0_{3\times 3} & \dots & 0_{3\times 3} & 0_{3\times 3} & \dots & J_{q}(x_N) \\
\end{array}
\right]}_{\underline{J}(x)}
\underbrace{
\begin{bmatrix}
\dot{p}_1 \\
\vdots \\
\dot{p}_N \\
\omega_1 \\
\vdots \\
\omega_N \\
\end{bmatrix}}_{\underline{v}(t)} \notag \\
&= \underline{J}(x) \underline{v}(t), \label{eq:v_underline}
\end{align}
and thus, \eqref{eq:e_stack_deriv_a} can be written as:
\begin{equation}
\dot{e}(t) = \bar{\mathbb{F}}_p(x) \bar{D}^\tau(\mathcal{G}) \underline{J}(x)\underline{v}(t). \label{eq:e_stack_deriv}
\end{equation}

\subsection{Performance Functions}
The concepts and techniques of prescribed performance control (see Section \ref{subsec:ppc}) are adapted in this work in order to: a) achieve predefined transient and steady state response for the distance and orientation errors $e^p_k, e^q_k,\forall k \in \mathcal{M}$ as well as ii) avoid the violation of the collision and connectivity constraints between neighboring agents, as presented in Section \ref{sec:prob_formulation}. The mathematical expressions of prescribed performance are given by the inequality objectives: 
\begin{subequations} \label{eq:ppc ineq}
\begin{align}
 -C_{k,\text{col}} \rho^p_k(t) &< e^p_k(t) < C_{k,\text{con}} \rho^p_k(t), \\
 -\rho^q_k(t) &< e^q_{k_n}(t) < \rho^q_k(t),
\end{align}
\end{subequations}
$\forall k \in \mathcal{M},n\in\{1,2,3\}$, where 
\begin{align*}
\rho^p_k(t) &= (1 - \dfrac{\rho^p_{k,\infty}} {\max\{C_{k,\text{con}}, C_{k,\text{col}} \}} )e^{-l^p_k t} \notag \\
&\hspace{+30mm}+ \dfrac{\rho^p_{k,\infty}} {\max\{C_{k,\text{con}}, C_{k,\text{col}} \}}, \\
\rho^q_k(t) &= (\rho^q_{k,0} - \rho^q_{k,\infty})e^{-l^q_k t} + \rho^q_{k,\infty}, 
\end{align*}
are designer-specified, smooth, bounded, and decreasing functions of time, where $l^p_k, l^q_k, \rho^p_{k,\infty}, \rho^q_{k,\infty}\in\mathbb{R}_{>0}, \forall k \in \mathcal{M}$, incorporate the desired transient and steady state performance specifications respectively, as presented in Section \ref{subsec:ppc}, and $C_{k,\text{col}}$, $C_{k,\text{con}}\in\mathbb{R}_{>0},\forall k \in \mathcal{M}$, are associated with the collision and connectivity constraints. In particular, we select
\begin{subequations} \label{eq:C_k}
\begin{align}
C_{k,\text{col}} &= d^2_k - d^2_{k,\text{col}}, \\
C_{k,\text{con}} &= d^2_{k,\text{con}} - d^2_k ,
\end{align}
\end{subequations}
$\forall k \in \mathcal{M}$, which, since the desired formation is compatible with the collision and connectivity constraints (i.e., $d_{k,\text{col}}<d_{k,\text{des}}<d_{k,\text{con}}, \forall k \in \mathcal{M}$), ensures that $C_{k,\text{col}},C_{k,\text{con}}\in\mathbb{R}_{>0},\forall k \in \mathcal{M}$ and consequently, in view of \eqref{eq:at t=0}, 
that:
\begin{subequations} \label{eq:ppc_time_0}
\begin{align}
-C_{k,\text{col}} \rho^p_k(0) < e^p_k(0) <\rho^p_k(0) C_{k,\text{con}}, 
\end{align}
$\forall k \in \mathcal{M}$. Moreover, by choosing 
\begin{equation}
\rho^q_{k,0} = \rho^q_k(0) > \max\limits_{n\in\{1,2,3\}}\lvert e^q_{k_n}(0) \rvert, \label{eq:rho_q_0}
\end{equation}
it is also guaranteed that:
\begin{align}
-\rho^q_k(0) < e^q_{k_n}(0) < \rho^q_k(0),
\end{align}
\end{subequations}
$\forall k \in \mathcal{M},n\in\{1,2,3\}$. Hence, if we guarantee prescribed performance via \eqref{eq:ppc ineq}, by employing the decreasing property of $\rho^p_k(t),\rho^q_k(t),\forall k \in \mathcal{M}$, we obtain:
\begin{align*}
-C_{k,\text{col}} &< e^p_k(t) < C_{k,\text{con}}, \\
-\rho^q_k(t) &< e^q_{k_n}(t) < \rho^q_k(t),
\end{align*}
and, consequently, owing to \eqref{eq:C_k}:
\begin{align*}
d_{k,\text{col}} < \lVert p_{\ell_k}(t)-p_{m_k}(t)\rVert < d_{k,\text{con}},
\end{align*}
$\forall k \in \mathcal{M}, t \in \mathbb{R}_{\geq 0}$, providing, therefore, a solution to problem \ref{problem}.

In the sequel, we propose a decentralized control protocol that does not incorporate any information on the agents' dynamic model and guarantees \eqref{eq:ppc ineq} for all $t\in\mathbb{R}_{\geq 0}$.

\subsection{Control Design}
Given the errors $e^p(t), e^q(t)$ defined in Section \ref{subsec:error derivation}: 

\textbf{Step I-a}: Select the corresponding functions $\rho^p_k(t), \rho^q_k(t)$ and positive parameters $C_{k,\text{con}}, C_{k,\text{col}}, k \in \mathcal{M}$, following \eqref{eq:ppc ineq}, \eqref{eq:rho_q_0}, and \eqref{eq:C_k}, respectively, in order to incorporate the desired transient and steady state performance specifications as well as the collision and connectivity constraints, and define the normalized errors $\xi_k^p:\mathbb{R}_{\geq 0} \to \mathbb{R},  \xi^q_k=[\xi_{k_1}^q,\xi^q_{k_2},\xi_{k_3}^q]^\tau:\mathbb{R}_{\geq 0} \to \mathbb{R}^3$:
\begin{subequations} \label{eq:ksi_k}
\begin{align}
&\hspace{-2mm}\xi^p_k(t) = (\rho^p_k(t))^{-1}e^p_k(t)  \\
&\hspace{-2mm}\xi^q_k(t) =  (\rho^q_k(t))^{-1} e^q_k(t), 
\end{align}
\end{subequations}
$\forall k \in \mathcal{M}$, as well as the stack vector forms 
\begin{align} 
\xi^p(t) &= [\xi^p_1(t),\dots,\xi^p_M(t)]^\tau = (\rho^p(t))^{-1}e^p(t), \notag \\ 
\xi^q(t) &= [(\xi^q_1(t))^\tau,\dots,(\xi^q_M(t))^\tau]^\tau = (\rho^q(t))^{-1}e^q(t), \notag \\ 
\xi(t) &= [(\xi^p(t))^\tau,(\xi^q(t))^\tau]^\tau = (\rho(t))^{-1}e(t)\in\mathbb{R}^{4M}, \label{eq:ksi_all}
\end{align}
where 
\begin{align*}
\rho^p(t) &= \text{diag}\{[\rho^p_k(t)]_{k \in \mathcal{M}}\}\in\mathbb{R}^{M\times M}, \\
\rho^q(t) &= \text{diag}\{[\rho^q_k(t) I_3]_{k \in \mathcal{M}}\}\in\mathbb{R}^{3M\times 3M}, \\
\rho(t) &= \text{diag}\{\rho^p(t), \rho^q(t) \}\in\mathbb{R}^{4M\times 4M}.
\end{align*}

\textbf{Step I-b}: Define the transformed errors $\varepsilon^p_k:\mathbb{R} \to \mathbb{R}, \varepsilon^q_k:\mathbb{R}^{3} \to \mathbb{R}^{3}$ and the signals $r^p_k:\mathbb{R} \to \mathbb{R}, r^q_k:\mathbb{R}^3 \to \mathbb{R}^{3\times3}$ as 
\begin{subequations} \label{eq:epsilon_p_and_phi_k}
\begin{align}
\varepsilon^p_k(\xi_k^p) &= \ln\left(\left(1 + \dfrac{\xi^p_k}{C_{k,\text{col}}}\right)\left(1 - \dfrac{\xi^p_k}{C_{k,\text{con}}}\right)^{-1} \right), \label{eq:epsilon_p_k}\\
\varepsilon^q_k(\xi_k^q) &= \left[\ln\left(\dfrac{1 + \xi^q_{k_1}}{1 - \xi^q_{k_1}}\right),\ln\left(\dfrac{1 + \xi^q_{k_2}}{1 - \xi^q_{k_2}}\right),\ln\left(\dfrac{1 + \xi^q_{k_3}}{1 - \xi^q_{k_3}}\right) \right]^\tau, \label{eq:epsilon_q_k} 
\end{align}
\end{subequations}
\begin{align}
&\hspace{-16mm}r^p_k(\xi^p_k) = \dfrac{\partial \varepsilon^p_k(\xi_k^p)}{\partial \xi_k^p} =  \dfrac{C_{k,\text{col}}+C_{k,\text{con}}}{(C_{k,\text{col}}+\xi_k^p)(C_{k,\text{con}}-\xi_k^p)},  \notag \\ 
&\hspace{-16mm}r^q_k(\xi^q_k) = \dfrac{\partial \varepsilon^q_k(\xi_k^q)}{\partial \xi_k^q} = \text{diag}\left\{\left[r^q_{k_n}(\xi^q_{k_n})\right]_{n\in\{1,2,3\}}\right\} \notag \\
&\hspace{-7mm}= \text{diag}\left\{\left[\dfrac{2}{1-(\xi_{k_n}^q)^2}\right]_{n\in\{1,2,3\}} \right\}, \notag 
\end{align}

and design the decentralized reference velocity vector for each agent $v_{i,\text{des}} = [\dot{p}^\tau_{i,\text{des}}, \omega^\tau_{i,\text{des}}]^\tau  :\mathbb{R}^{4M} \times\mathbb{R}_{\geq 0} \to \mathbb{R}^6$ as:
\begin{align}
 v_{i,\text{des}}(\xi,t) &= \notag \\
&\hspace{-14mm}-J^{-1}_i(x_i) \begin{bmatrix}
 \sum\limits_{j\in\mathcal{N}_i(0)} (\rho^p_{k_{ij}}(t))^{-1}r^p_{k_{ij}}(\xi^p_{k_{ij}})\varepsilon^p_{k_{ij}}(\xi^p_{k_{ij}})p_{i,j}(t) \\
 \sum\limits_{j\in\mathcal{N}_i(0)} (\rho^q_{k_{ij}}(t))^{-1}r^q_{k_{ij}}(\xi^q_{k_{ij}})\varepsilon^q_{k_{ij}}(\xi^q_{k_{ij}}) \label{eq:vel_i_des}
\end{bmatrix}  
\end{align}

where $k_{ij}\in \mathcal{M}$ is the edge of agents $i,j\in\mathcal{N}_i(0)$, i.e., $\{\ell_{k_{ij}},m_{k_{ij}}\}\in\mathcal{E}$ and $\ell_{k_{ij}}=i,m_{k_{ij}}=j$. The desired velocities \eqref{eq:vel_i_des} can be written in vector form: 
\begin{align}
\underline{v}_{\text{des}}(\xi,t) &= \begin{bmatrix} \dot{p}_{\text{des}}(\xi^p,t)\\\omega_{\text{des}}(\xi^q,t)  \end{bmatrix} \notag \\ 	
&= -\underline{J}^{-1}(x)\bar{D}(\mathcal{G})\bar{\mathbb{F}}^\tau_p(x)r(\xi)(\rho(t))^{-1}\varepsilon(\xi), \label{eq:vel_des}
\end{align}
	
where $\dot{p}_{\text{des}} = [\dot{p}^\tau_{1,\text{des}},\dots,\dot{p}^\tau_{N,\text{des}}]^\tau, \omega_{\text{des}} = [\omega^\tau_{1,\text{des}},\dots, \\ \omega^\tau_{N,\text{des}}]^\tau\in\mathbb{R}^{3N}, \varepsilon = [(\varepsilon^p)^\tau,(\varepsilon^q)^\tau]^\tau = [\varepsilon^p_1,\dots,\varepsilon^p_M,(\varepsilon^q_1)^\tau, \\ \dots, (\varepsilon^q_M)^\tau]^\tau\in\mathbb{R}^{4M}$ and $\underline{J}(x), \bar{D}(\mathcal{G}),\bar{\mathbb{F}}_p$ as they were defined in \eqref{eq:bar_F_bar_D} and \eqref{eq:v_underline}, respectively. Moreover,  
\begin{equation*}
r = \begin{bmatrix}
r^p & 0_{M\times 3M} \\ 
0_{3M\times M} & r^q
\end{bmatrix} \in\mathbb{R}^{4M\times 4M},
\end{equation*}
$r^p = \text{diag}\{[r^p_k]_{k \in \mathcal{M}}\}\in\mathbb{R}^{M\times M}$ and $r^q =\text{diag}\{[r^q_k]_{k \in \mathcal{M}}\}\in\mathbb{R}^{3M\times 3M}$. It should be noted that $\underline{J}^{-1}(x)$ is always well-defined due to Assumption 1.

\textbf{Step II-a}: Define the velocity errors $e^v:\mathbb{R}^{4M}\times\mathbb{R}_{\geq 0} \to \mathbb{R}^{6N}$, with $e^v(\xi,t) = [(e^v_1)^\tau(\xi,t),\dots,(e^v_N)^\tau(\xi,t)]^\tau = v(t) - v_{\text{des}}(\xi,t)$\footnote{Notice the difference between $\underline{v}_{\text{des}} = [\dot{p}^\tau_{\text{des}},\omega^\tau_{\text{des}}]^\tau$  and $v_{\text{des}} = [\dot{p}^\tau_{1,\text{des}},\omega^\tau_{1,\text{des}},\dots,\dot{p}^\tau_{N,\text{des}},\omega^\tau_{N,\text{des}}]^\tau$.}, where  $e^v_i(\xi,t)= [e^v_{i_1}(\xi,t),\dots,e^v_{i_6}(\xi,t)]^\tau=[\dot{p}^\tau_i(t)-\dot{p}^\tau_{i,\text{des}}(\xi^p,t),\omega_i^\tau(t)-\omega^\tau_{i,\text{des}}(\xi^q,t)]^\tau = v_i(t)-v_{i,\text{des}}(\xi,t),i\in\mathcal{V}$, and select the corresponding performance functions $\rho^v_{i_m}:\mathbb{R}_{\geq 0} \to \mathbb{R}_{>0}$, with $\rho^v_{i_m}(t) = (\rho^v_{i_m,0}-\rho^v_{i_m,\infty})e^{-l^v_{i_m}t} + \rho^v_{i_m,\infty}$ and $\rho^v_{i_m,0}=\rho^v_{i_m}(0)> \lvert e^v_{i_m}(0) \rvert,l^v_{i_m},\rho^v_{i_m,\infty}\in\mathbb{R}_{>0}, \rho^v_{i_m,\infty}<\rho^v_{i_m,0}, \forall i\in\mathcal{V},m\in\{1,\dots,6\}$. Moreover, define the normalized velocity errors $\xi_i^v=[\xi_{i_1}^v,\dots,\xi_{i_6}^v]^\tau:\mathbb{R}^{4M}\times\mathbb{R}_{\geq 0} \to \mathbb{R}^{6}$:
\begin{equation}
\xi_i^v(\xi,t) = (\rho^v_i(t))^{-1}e^v_i(\xi,t), \notag \\\label{eq:ksi_i_v}
\end{equation}
with $\rho^v_i(t) = \text{diag}\{[\rho^v_{i_m}(t)]_{m\in\{1,\dots,6\}}\}\in\mathbb{R}^{6\times6}$, which is written in vector form as:
\begin{align}
\xi^v(\xi,t) &= [(\xi_1^v(\xi,t))^\tau,\dots,(\xi_N^v(\xi,t))^\tau]^\tau \notag \\
&= (\rho^v(t))^{-1}e^v(\xi,t)\in\mathbb{R}^{6N}, \label{eq:ksi_v}
\end{align}

with $\rho^v(t) = \text{diag}\left\{\left[\rho^v_i(t)\right]_{i\in\mathcal{V}}\right\}\in\mathbb{R}^{6N\times6N}$. 

\textbf{Step II-b}: Define the transformed velocity errors $\varepsilon_i^v:\mathbb{R}^{6} \to \mathbb{R}^{6}$ and the signals $r^v_i:\mathbb{R}^{6} \to \mathbb{R}^{6\times 6}$ as:
\begin{subequations}
\begin{align}
&\varepsilon_i^v(\xi_i^v) = \left[\ln\left(\dfrac{1+\xi^v_{i_1}}{1-\xi^v_{i_1}}\right),\cdots,\ln\left(\dfrac{1+\xi^v_{i_6}}{1-\xi^v_{i_6}}\right)\right]^\tau, \label{eq:epsilon_v}\\
& r^v_i(\xi_i^v) = \dfrac{\partial \varepsilon_i^v(\xi_i^v)}{\partial \xi_i^v} = \text{diag}\{\left[r^v_{i_m}(\xi^v_{i_m})\right]_{m\in\{1,\dots,6\}} \}\notag \\
&\hspace{9.5mm}= \text{diag}\left\{\left[ \dfrac{2}{(1-(\xi^v_{i_m})^2)} \right]_{m\in\{1,\dots,6\}} \right\},
\end{align}
\end{subequations}

and design the decentralized control protocol for each agent $i\in\mathcal{V}$ as $u_i:\mathbb{R}^{6}\times\mathbb{R}_{\geq 0} \to \mathbb{R}^{6}$:
\begin{equation}
u_i(\xi_i^v,t) = -\gamma_i(\rho_i^v(t))^{-1}r_i^v(\xi_i^v)\varepsilon_i^v(\xi_i^v), \label{eq:u_i}
\end{equation}
with $\gamma_i \in\mathbb{R}_{>0}, \forall i\in\mathcal{V}$, which can be written in vector form as: 
\begin{equation}
u(\xi^v,t) = -\Gamma(\rho^v(t))^{-1}r^v(\xi^v)\varepsilon^v(\xi^v), \label{eq:u}
\end{equation}
where $\Gamma=\text{diag}\{[\gamma_iI_6]_{i\in\mathcal{V}}  \}\in\mathbb{R}^{6N\times 6N}$, $\varepsilon^v = [(\varepsilon_1^v)^\tau,\dots, \\ (\varepsilon_N^v)^\tau]^\tau\in\mathbb{R}^{6N}$ and $r^v=\text{diag}\{[r_i^v]_{i\in\mathcal{V}}\}\in\mathbb{R}^{6N\times6N}$.

\begin{remark} Note that the selection of $C_{k,\text{col}}, C_{k,\text{con}}$ according to \eqref{eq:C_k} and of $\rho^q_{k}(t),\rho^v_{i_m}(t)$ such that
$\rho^q_{k,0}=\rho^q_k(0)>\max\limits_{n\in\{1,2,3\}}\lvert e^q_{k_n}(0)\rvert, \rho^v_{i_m,0}=\rho^v_{i_m}(0)>\lvert e^v_{i_m}(0)\rvert$ along with \eqref{eq:at t=0}, guarantee that $\xi^p_{k}(0)\in(C_{k,\text{col}},C_{k,\text{con}})$, $\xi^q_{k_n}(0)\in(-1,1)$, $\xi^v_{i_m}(\xi(0),0)\in(-1,1)$, $\forall k \in \mathcal{M},n\in\{1,2,3\},m\in\{1,\dots,6\}, i\in\mathcal{V}$.
The prescribed performance control technique enforces these normalized 
errors $\xi^p_{k}(t), \xi^q_{k_n}(t)$ and $\xi^v_{i_m}(t)$ to remain strictly within the sets $(
-C_{k,\text{col}},C_{k,\text{con}}), (-1,1)$, and
$(-1,1)$, respectively, $\forall k \in \mathcal{M},n\in\{1,2,3\},m\in\{1,\dots,6\}, i\in\mathcal{V},t\geq0$, guaranteeing thus a solution to Problem \ref{problem}. It can be verified that this can be achieved by maintaining the boundedness of the modulated
errors $\varepsilon^{p}(\xi^{p}(t)), \varepsilon^{q}(\xi^{q}(t))$ and $\varepsilon^{v}%
(\xi^{v}(t), \forall t\geq0$.
\end{remark}

\begin{remark}
Notice by \eqref{eq:vel_i_des} and \eqref{eq:u_i} that the proposed control protocols are distributed in the sense
that each agent uses only local information to calculate its
own signal. In that respect, regarding every edge $k_{ij}$, with $\{\ell_{k_{ij}},m_{k_{ij}}\}=\{i,j\}$, the parameters $\rho^p_{k_{ij},\infty}, \rho^q_{k_{ij},\infty}, l^p_{k_{ij}}, l^q_{k_{ij}}$, as well as the sensing radii $s_j,\forall j\in \mathcal{N}_i(0)$, which are needed for the calculation of the performance functions $\rho^p_{k_{ij}}, \rho^q_{k_{ij}}$, can be transmitted off-line to each agent $i\in\mathcal{V}$. It should also be noted that the proposed
control protocol \eqref{eq:u_i} depends exclusively on
the velocity of each agent and not on the velocity of its neighbors.
Moreover, the proposed control law does not incorporate any prior knowledge of
the model nonlinearities/disturbances, enhancing thus its robustness. Furthermore, the proposed
methodology results in a low complexity. Notice that no hard
calculations (neither analytic nor numerical) are required to output the
proposed control signal.
\end{remark}

\begin{remark}
Regarding the construction of the performance functions, we stress that the
desired performance specifications concerning the transient and steady state
response as well as the collision and connectivity constraints are introduced
in the proposed control schemes via $\rho^p_k(t), \rho^q_k(t)$ and
$C_{k,\text{col}}, C_{k,\text{con}} $, $k \in \mathcal{M}$.
In addition, the velocity performance functions $\rho^v_{i_m}(t)
$, impose prescribed performance on the velocity errors
$e^v_i=v_i-v_{i,\text{des}}$, $i\in\mathcal{V}$. In this respect, notice that
$v_{i,\text{des}}$ acts as a reference signal for the corresponding
velocities $v_{i}$, $i\in\mathcal{V}$. However, it should be stressed that
although such performance specifications are not required (only the
neighborhood position and orientation errors need to satisfy predefined transient and steady
state performance specifications), their selection affects both the evolution
of the errors within the corresponding performance envelopes as well
as the control input characteristics (magnitude and rate). Nevertheless, the
only hard constraint attached to their definition is related to their initial
values. Specifically, $\rho^q_{k,0}=\rho^q_k(0)>\max\limits_{n\in\{1,2,3\}}\lvert e^q_{k_n}(0)\rvert, \rho^v_{i_m,0}=\rho^v_{i_m}(0)>\lvert e^v_{i_m}(0)\rvert$, $\forall k \in \mathcal{M},n\in\{1,2,3\},m\in\{1,\dots,6\}, i\in\mathcal{V}$.
\end{remark}

\subsection{Stability Analysis}
The main results of this work are summarized in the following theorem.

\begin{thm}
Consider a system of $N$ rigid bodies aiming at establishing a formation described by the desired distances $d_{k,\text{des}}$ and orientation angles $q_{k,\text{des}}, k\in\mathcal{M}$, while satisfying the collision and connectivity constraints between neighboring agents, represented by $d_{k,\text{col}}$ and $d_{k,\text{con}}$, respectively, with $d_{k,\text{col}} < d_{k,\text{des}} < d_{k,\text{con}}, k\in\mathcal{M}$. Then, under Assumptions \ref{as:J}, \ref{assump:basic_assumption}, the decentralized control protocol \eqref{eq:ksi_k}-\eqref{eq:u} guarantees: 
\begin{align}
-C_{k,\text{col}} \rho^p_k(t) &< e^p_k(t) < C_{k,\text{con}} \rho^p_k(t), \notag\\
-\rho^q_k(t) &< e^q_{k_n}(t) < \rho^q_k(t), \notag
\end{align}
$\forall k\in\mathcal{M},n\in\{1,2,3\},t\geq 0$, as well as the boundedness of all closed loop signals.
\end{thm}
\begin{pf}
By differentiating \eqref{eq:ksi_all} and \eqref{eq:ksi_v} with respect to time, we obtain:
\begin{align}
\dot{\xi}(\xi,t) &= (\rho(t))^{-1} \left[\dot{e}(t) - \dot{\rho}(t)\xi \right], \notag\\
\dot{\xi}^v(\xi,\xi^v,t) &= (\rho^v(t))^{-1}\left[ \dot{e}^v(\xi,t) - \dot{\rho}^v(t)\xi^v \right], \notag
\end{align}
which, by substituting \eqref{eq:e_stack_deriv} and \eqref{eq:system}, becomes:
\begin{align}
&\hspace{-2mm}\dot{\xi}(\xi,t) =  (\rho(t))^{-1}\left[\bar{\mathbb{F}}_p(x)\bar{D}^\tau(\mathcal{G})\underline{J}(x)\underline{v}(t) -  \dot{\rho}(t)\xi\right], \notag\\
&\hspace{-2mm}\dot{\xi}^v(\xi,\xi^v,t) = (\rho^v(t))^{-1}\left\{\bar{M}^{-1}(x)\left[u - \bar{C}(x,\dot{x})v - \bar{g}(x)  \right. \right. \notag\\
& \left. \left. \hspace{18mm}- \bar{w}(x,\dot{x},t)\right] - \dot{v}_{\text{des}}(\xi,t) -  \dot{\rho}^v(t)\xi^v\right\}. \notag
\end{align}
By employing \eqref{eq:vel_des}, \eqref{eq:u} as well as the fact that $v(t) = e^v(\xi,t) + v_{\text{des}}(\xi,t)= \rho^v(t)\xi^v(\xi,t) + v_{\text{des}}(\xi,t)$ from \eqref{eq:ksi_v}, the following is obtained:
\begin{subequations} \label{eq:ksi_dot_3}
\begin{align}
&\hspace{-2mm}\dot{\xi} =  h(\xi,t) \notag\\
&\hspace{-2mm}= -(\rho(t))^{-1}P(x)r(\xi)(\rho(t))^{-1}\varepsilon(\xi) - (\rho(t))^{-1}\dot{\rho}(t)\xi \notag\\ 
&\hspace{2mm} + (\rho(t))^{-1}\bar{\mathbb{F}}_p(x)\bar{D}^\tau(\mathcal{G})\underline{J}(x)\rho^v(t)\xi^v(\xi,t),  \label{eq:ksi_dot_3_p_phi}
\end{align}
\begin{align}
&\hspace{-2mm}\dot{\xi}^v = h^v(\xi,\xi^v,t) \notag\\
&\hspace{-2mm}= -(\rho^v(t))^{-1}\bar{M}^{-1}(x)\Gamma(\rho^v(t))^{-1}r^v(\xi^v)\varepsilon^v(\xi^v) \notag\\
&\hspace{-2mm}- (\rho(t))^{-1}\left\{\bar{M}^{-1}(x)\left[\bar{C}(x,\dot{x})(\rho^v(t)\xi^v(\xi,t) + v_{\text{des}}(\xi,t))  \right. \right. \notag\\
&\hspace{-2mm}\left. \left. + \bar{g}(x)+ \bar{w}(x,\dot{x},t)\right] + \dot{v}_{\text{des}}(\xi,t) +  \dot{\rho}^v(t)\xi^v\right\},   \label{eq:ksi_dot_3_v}
\end{align}
\end{subequations}
where $P(x) = \bar{\mathbb{F}}_p(x)\bar{D}^\tau(\mathcal{G})\bar{D}(\mathcal{G})\bar{\mathbb{F}}^\tau_p(x)$.

By defining $\bar{\xi}=[\xi^\tau,(\xi^v)^\tau]^\tau\in\mathbb{R}^{4M+6N}$, the closed loop system of \eqref{eq:ksi_dot_3} can be written in compact form as:
\begin{equation}
\dot{\bar{\xi}} = \bar{h}(t,\bar{\xi}) = \begin{bmatrix}h(\xi,t)\\h^v(\xi,\xi^v,t) \end{bmatrix}. \label{eq:ksi_dot_compact}
\end{equation}
Let us also define the open set $\Omega_{\bar{\xi}} = \Omega_{\xi^p}\times\Omega_{\xi^q}\times\Omega_{\xi^v}$, with 
\begin{align}
\Omega_{\xi^p} &= (-C_{1,\text{col}},C_{1,\text{con}})\times\cdots\times(-C_{M,\text{col}},C_{M,\text{con}}), \notag\\
\Omega_{\xi^q} & = (-1,1)^{3M}, \notag \\ 
\Omega_{\xi^v} & = (-1,1)^{6N}. \notag
\end{align}
In what follows, we proceed in two phases. First, the existence of a unique maximal solution $\bar{\xi}(t)$ of \eqref{eq:ksi_dot_compact} over the set $\Omega_{\bar{\xi}}$ for a time interval $[0,\tau_{\max})$ is ensured (i.e., $\bar{\xi}(t)\in\Omega_{\bar{\xi}},\forall t\in[0,\tau_{\max})$). Then, we prove that the proposed control scheme \eqref{eq:vel_des} and \eqref{eq:u} guarantees, for all $t\in[0,\tau_{\max})$,  the boundedness of all closed loop signals, as well as that $\bar{\xi}(t)$ remains strictly within a compact subset of $\Omega_{\bar{\xi}}$, which leads by contradiction to $\tau_{\max} = +\infty$.

\subsubsection{\textbf{Phase A}:} By selecting the parameters $C_{k,\text{col}},C_{k,\text{con}},k\in\mathcal{M}$, according to \eqref{eq:C_k}, we guarantee that the set $\Omega_{\bar{\xi}}$ is nonempty and open. Moreover, as shown in \eqref{eq:ppc_time_0}, we guarantee that $\xi^p(0)\in\Omega_{\xi^p}$ and $\xi^q(0)\in\Omega_{\xi^q}$. In addition, by selecting $\rho^v_{i_m}(0)>\lvert e^v_{i_m}(0) \rvert, \forall i\in\mathcal{V},m\in\{1,\dots,6\}$, we also guarantee that $\xi^v(0)\in\Omega_{\xi^v}$. Hence, $\bar{\xi}(0)\in\Omega_{\bar{\xi}}$. Furthermore, $\bar{h}$ is continuous on $t$ and locally Lipschitz on $\bar{\xi}$ over the set $\Omega_{\bar{\xi}}$. Therefore, according to Theorem \ref{thm:dynamical systems} in Section \ref{subsec:dynamical systems}, there exists a maximal solution $\bar{\xi}(t)$ of \eqref{eq:ksi_dot_compact} on the time interval $[0,\tau_{\max})$ such that $\bar{\xi}(t)\in\Omega_{\bar{\xi}}, \forall t\in[0,\tau_{\max})$.

\subsubsection{\textbf{Phase B}:} We have proven in Phase A that $\bar{\xi}(t)\in\Omega_{\bar{\xi}}, \forall t\in[0,\tau_{\max})$ and more specifically, that
\begin{subequations} \label{eq:ksi_bounded}
\begin{align}
\xi^p_k(t) &= \dfrac{e^p_k(t)}{\rho^p_k(t)}\in(-C_{k,\text{col}},C_{k,\text{con}}), \label{eq:ksi_p_bounded}\\
\xi^q_{k_n}(t) &= \dfrac{e^q_{k_n}(t)}{\rho^p_k(t)}\in(-1,1),\\
\xi^v_{i_m}(t) &= \dfrac{e^v_{i_m}(t)}{\rho^v_i(t)}\in(-1,1), \label{eq:ksi_v_bounded} 
\end{align}
\end{subequations}
$\forall k\in\mathcal{M},n\in\{1,2,3\},m\in\{1,\dots,6\},i\in\mathcal{V}$, from which we conclude that $e^p_k(t), e^q_{k_n}(t)$ and $e^v_{i_m}(t)$ are bounded by $\max\{C_{k,\text{col}},C_{k,\text{con}}\}, \rho^q_k(t)$ and  $\rho^v_{i_m}(t)$, respectively, $\forall t\in[0,\tau_{\max})$. Furthermore, the error vector $\varepsilon(\xi)$, as given in \eqref{eq:vel_des}, is well defined $\forall t\in[0,\tau_{\max})$. Therefore, consider the positive definite and radially unbounded function $V_1:\mathbb{R}^{4M} \to \mathbb{R}_{\geq 0}$, with $V_1(\varepsilon) = \tfrac{1}{2}\varepsilon^\tau\varepsilon$. Time differentiation of $V_1$ yields $\dot{V}_1 = \varepsilon^\tau r(\xi)\dot{\xi}$, which, after substituting \eqref{eq:ksi_dot_3_p_phi}, becomes 
\begin{align}
&\hspace{-3mm}\dot{V}_1 = -\varepsilon^\tau r(\xi)(\rho(t))^{-1}P(x)r(\xi)(\rho(t))^{-1}\varepsilon \notag\\
&\hspace{-3mm}- \varepsilon^\tau r(\xi)(\rho(t))^{-1}\left[\dot{\rho}(t)\xi - \bar{\mathbb{F}}_p(x)\bar{D}^\tau(\mathcal{G})\underline{J}(x)\rho^v(t)\xi^v \right]. \notag
\end{align}
Note that: 1) $\dot{\rho}(t),\rho^v(t), \bar{D}(\mathcal{G})$ are bounded by construction, 2) $\underline{J}$ and $\xi^v, p,q$ are bounded $\forall t\in[0,\tau_{\max})$ owing to Assumption \ref{as:J} and \eqref{eq:ksi_bounded}, respectively, and hence $\bar{\mathbb{F}}_p(x)$ is also bounded $\forall t\in[0,\tau_{\max})$ due to its continuity. Therefore, by also exploiting the fact that $\rho(t),r(\xi)$ are diagonal, $\dot{V}_1$ becomes
\begin{align}
\dot{V}_1 & \leq -((\rho(t))^{-1}r(\xi)\varepsilon)^\tau P(x) (r(\xi)(\rho(t))^{-1}\varepsilon) \notag\\
&\hspace{+40mm}+ \lVert (\rho(t))^{-1} r(\xi)\varepsilon\rVert\bar{B}_1, \notag
\end{align}
where $\bar{B}_1$ is a positive constant, independent of $\tau_{\max}$, satisfying 
\begin{equation}
\lVert\dot{\rho}(t)\xi - \bar{\mathbb{F}}_p(x)\bar{D}^\tau(\mathcal{G})\underline{J}(x)\rho^v(t)\xi^v \rVert \leq \bar{B}_1,  \label{eq:bar B_1}
\end{equation}
By invoking Lemma \ref{lem: P > 0} from Appendix A, $\dot{V}_1$ becomes
\begin{align}
&\dot{V}_1 \leq  -\lambda_{\min}(P)\lVert (\rho(t))^{-1}r(\xi)\varepsilon)\rVert^2 + \lVert (\rho(t))^{-1} r(\xi)\varepsilon\rVert\bar{B}_1 \notag \\
&\leq  -\lVert (\rho(t))^{-1}r(\xi)\varepsilon)\rVert \left[   \lambda_{\min}(P)\lVert (\rho(t))^{-1}r(\xi)\varepsilon)\rVert- \bar{B}_1 \right], \notag 
\end{align}
with $\lambda_{\min}(P)>0$. Therefore, $\dot{V}_1 < 0$ when $\lVert (\rho(t))^{-1}r(\xi)\varepsilon)\rVert \\ > \dfrac{\bar{B}_1}{\lambda_{\min}(P)}$. By using the definitions of $r(\xi)$ and $\rho(t)$ as well as their positive definiteness $\forall t\in[0,\tau_{\max})$, the last inequality can be shown to be equivalent to $\lVert\varepsilon \rVert > \dfrac{\bar{B}_1 \tilde{r}}{\lambda_{\min}(P)}$, where $\tilde{r} = \max\{\max\limits_{k\in\mathcal{M}}\{C_{k,\text{col}}+C_{k,\text{con}}\}, \max\limits_{k\in\mathcal{M}}\{\rho^q_{k,0}\} \}$. Therefore, we conclude that
\begin{equation}
\lVert \varepsilon(\xi(t)) \rVert \leq \bar{\varepsilon} = \max \left\{ \varepsilon(\xi(0)), \dfrac{\bar{B}_1 \tilde{r}}{\lambda_{\min}(P)} \right\}, 
\end{equation} 
$\forall t\in[0,\tau_{\max})$. Furthermore, from \eqref{eq:epsilon_p_and_phi_k}, by taking the inverse logarithm function, we obtain:
\begin{subequations} \label{eq:e_bar_p_q}
\begin{align}
 &\hspace{-3.5mm}-C_{k,\text{col}} < \dfrac{e^{-\bar{\varepsilon}}-1}{e^{-\bar{\varepsilon}}+1}C_{k,\text{col}} = \xi^p_{k,\min}\leq \xi^p_k(t) \leq \xi^p_{k,\max}\notag\\ &\hspace{+35mm}=\dfrac{e^{\bar{\varepsilon}}-1}{e^{\bar{\varepsilon}}+1}C_{k,\text{con}} <  C_{k,\text{con}}, \label{eq:e_bar_p} \\
 &\hspace{-3.9mm}-1 < \dfrac{e^{-\bar{\varepsilon}}-1}{e^{-\bar{\varepsilon}}+1}=\xi^q_{\min} \leq \xi^q_{k_n}(t) \leq \xi^q_{\max}= \dfrac{e^{\bar{\varepsilon}}-1}{e^{\bar{\varepsilon}}+1} < 1,  \label{eq:e_bar_q}
\end{align}
\end{subequations}
$\forall t\in[0,\tau_{\max}),k\in\mathcal{M},n\in\{1,2,3\}$. Thus, the reference velocity vector $\underline{v}_{\text{des}}(\xi,t)$, as designed in \eqref{eq:vel_des}, remains bounded $\forall t\in[0,\tau_{\max})$. Moreover, since $v(t)=\rho^v(t)\xi^v(\xi,t)+v_{\text{des}}(\xi,t)$, we also conclude the boundedness of $v(t),\forall t\in[0,\tau_{\max})$. Finally, differentiating $v_{\text{des}}$ with respect to time, substituting \eqref{eq:ksi_dot_3_p_phi} and using \eqref{eq:e_bar_p_q}, the boundedness of $\dot{v}_{\text{des}}, \forall t\in[0,\tau_{\max})$, is deduced as well.

Applying the aforementioned line of proof, we consider the positive definite and radially unbounded function $V_2:\mathbb{R}^{6N} \to \mathbb{R}_{\geq 0}$, with $V_2(\varepsilon^v) = \tfrac{1}{2}(\varepsilon^v)^\tau\Gamma\varepsilon^v$, since the error vector $\varepsilon^v(\xi^v)$ is well defined $\forall t\in[0,\tau_{\max})$, due to \eqref{eq:ksi_v_bounded}. Time differentiation of $V_2$ yields $\dot{V}_2 = (\varepsilon^v)^\tau \Gamma r^v(\xi^v)\dot{\xi}^v$, which, after substituting \eqref{eq:ksi_dot_3_v}, becomes 
\begin{align}
&\dot{V}_2 = -(\varepsilon^v)^\tau\Gamma r^v(\xi^v)(\rho^v(t))^{-1}\bar{M}^{-1}(x)\Gamma(\rho^v(t))^{-1}r^v(\xi^v)\varepsilon^v \notag\\
&-(\varepsilon^v)^\tau r^v(\xi^v)(\rho^v(t))^{-1}\left\{ \bar{M}^{-1}(x)\left[\bar{g}(x) + \bar{w}(x,\dot{x},t) + \right. \right. \notag\\
&\left. \left. \bar{C}(x,\dot{x})(\rho^v(t)\xi^v(\xi,t) + v_{\text{des}}(\xi,t))\right] + \dot{v}_{\text{des}}(\xi,t) + \dot{\rho}^v(t)\xi^v\right\}. \notag
\end{align}
By exploiting the boundedness of $\xi^v$ and the positive definiteness and diagonality of $\Gamma, \rho^v(t), r^v(\xi^v),\forall t\in[0,\tau_{\max})$ due to \eqref{eq:ksi_v_bounded}, the boundedness of $\rho^v,\dot{\rho^v},v_{\text{des}},\dot{v}_{\text{des}}, \bar{w}(x,\dot{x},t)$, the continuity of $\bar{M}^{-1}, \bar{C}, \bar{g}$ and the positive definiteness of $\bar{M}^{-1}$,  $\dot{V}_2$ becomes 
\begin{align}
\dot{V}_2 \leq -\lambda_{\min}(\Gamma \bar{M}^{-1}\Gamma)\lVert (\rho^v(t))^{-1}r^v(\xi^v)\varepsilon^v(\xi^v) \rVert^2 + \notag\\
\lVert (\rho^v(t))^{-1}r^v(\xi^v)\varepsilon^v(\xi^v)\rVert \bar{B}_2 \notag,
\end{align}
where $\lambda_{\min}(\Gamma \bar{M}^{-1}\Gamma)>0$ and $\bar{B}_2$ is a positive constant, independent of $\tau_{\max}$, that satisfies 
\begin{align}
 \lVert \bar{M}^{-1}(x)\left(\bar{g}(x) + \bar{w}(x,\dot{x},t) + \bar{C}(x,\dot{x})(\rho^v(t)\xi^v(\xi,t) \right.  \notag\\
\left.   + v_{\text{des}}(\xi,t))\right) + \dot{v}_{\text{des}}(\xi,t) + \dot{\rho}^v(t)\xi^v \rVert \leq \bar{B}_2. \notag
\end{align}
Therefore, we conclude that $\dot{V}_2 < 0$ when $$\lVert (\rho^v(t))^{-1}r^v(\xi^v)\varepsilon^v(\xi^v) \rVert > \dfrac{\bar{B}_2}{\lambda_{\min}(\Gamma \bar{M}^{-1}\Gamma)},$$ which is equivalent to $\lVert \varepsilon^v \rVert > \dfrac{\bar{B}_2 \tilde{r}_v}{\lambda_{\min}(\Gamma \bar{M}^{-1}\Gamma)}$, with $\tilde{r}_v = \max \left\{\rho^v_{i_m,0}, i\in\mathcal{V},m\in\{1,\dots,6\}\right\}$. Hence, we conclude that:
\begin{align}
\lVert \varepsilon^v(\xi^v(\xi(t),t)) \rVert &\leq \bar{\varepsilon}^v \notag \\
&\hspace{-18mm}= \max \left\{\varepsilon^v(\xi^v(\xi(0),0)), \dfrac{\bar{B}_2 \tilde{r}_v}{ \lambda_{\min}(\Gamma \bar{M}^{-1}\Gamma)} \right\} \notag 
\end{align}
$\forall t\in[0,\tau_{\max})$. Furthermore, from \eqref{eq:epsilon_v}, we obtain:
\begin{align}
-1 < \dfrac{e^{-\bar{\varepsilon}^v}-1}{e^{-\bar{\varepsilon}^v}+1}=\xi^v_{\min} &\leq \xi^v_{i_m}(t) \notag \\ 
&\leq \xi^v_{\max}= \dfrac{e^{\bar{\varepsilon}^v}-1}{e^{\bar{\varepsilon}^v}+1} < 1, \label{eq:e_bar_v}
\end{align}
$\forall t\in[0,\tau_{\max}),m\in\{1,\dots,6\},i\in\mathcal{V}$, which leads to the boundedness of the decentralized control protocol \eqref{eq:u}.

Up to this point, what remains to be shown is that $\tau_{\max}$ can be extended to $\infty$. In this direction, notice by \eqref{eq:e_bar_p_q} and \eqref{eq:e_bar_v} that $\bar{\xi}(t)\in\Omega'_{\bar{\xi}} = \Omega'_{\xi^p}\times\Omega'_{\xi^q}\times\Omega'_{\xi^v}, \forall t\in[0,\tau_{\max})$, where: 
\begin{align}
\Omega'_{\xi^p} &= [\xi^p_{1,\min}, \xi^p_{1,\max}]\times\cdots\times[\xi^p_{M,\min}, \xi^p_{M,\max}], \notag \\
\Omega'_{\xi^q} &= [\xi^q_{\min}, \xi^q_{\max}]^{3M}, \notag \\
\Omega'_{\xi^v} &= [\xi^v_{\min}, \xi^v_{\max}]^{6N}, \notag 
\end{align}
are nonempty and compact subsets of $\Omega_{\xi^p},\Omega_{\xi^q}$ and $\Omega_{\xi^v}$, respectively. Hence, assuming that $\tau_{\max} < \infty$ and since $\Omega'_{\bar{\xi}}\subseteq\Omega_{\bar{\xi}}$, Proposition \ref{prop:dynamical systems} in Section \ref{subsec:dynamical systems} dictates the existence of a time instant $t'\in[0,\tau_{\max})$ such that $\bar{\xi}(t')\notin\Omega'_{\bar{\xi}}$, which is a contradiction. Therefore, $\tau_{\max} = \infty$. Thus, all closed loop signals remain bounded and moreover $\bar{\xi}(t)\in\Omega'_{\bar{\xi}}\subseteq\Omega_{\bar{\xi}}, \forall t\in\mathbb{R}_{\geq 0}$. Finally, multiplying \eqref{eq:e_bar_p} and \eqref{eq:e_bar_q} by $\rho^p_k(t)$ and $\rho^q_k(t)$, respectively, we also conclude: 
\begin{subequations}
\begin{align*}
-C_{k,\text{col}}\rho^p_k(t) &< e^p_k(t) < C_{k,\text{con}}\rho^p_k(t), \\
-\rho^q_k(t) &< e^q_{k_n}(t) < \rho^q_k(t), 
\end{align*}
\end{subequations}
$\forall k\in\mathcal{M},n\in\{1,2,3\},t\in\mathbb{R}_{\geq 0}$, which leads to the completion of the proof. 
\end{pf}

\begin{remark}
Notice that \eqref{eq:e_bar_p_q} and \eqref{eq:e_bar_v} hold no matter how large
the finite bounds $\bar{\varepsilon}$, $\bar{\varepsilon}^v$ are. Therefore, there is no need to render $\bar{\varepsilon}^v$ arbitrarily small by adopting
extreme values of the control gains $\gamma_i$.
In the same spirit, large uncertainties involved in the nonlinear model
\eqref{eq:system} can be compensated, as they affect only the
size of $\bar{\varepsilon}^{v}$ through $\bar{B}_{2}$, but leave unaltered the achieved stability
properties. Hence, the actual performance of the system becomes isolated against model
uncertainties, thus enhancing the robustness of the proposed control schemes.
\end{remark}

\begin{remark}\label{remark_5}
The transient and steady state performance of
the closed loop system is explicitly and solely determined by appropriately
selecting the parameters $l^p_k,l^q_k$, $\rho^p_{k,\infty}, \rho^q_{k,\infty}, \rho^p_{k,0}$ and $C_{k,\text{col}}$,
$C_{k,\text{con}}$, $k\in\mathcal{M}$. In that respect, the performance attributes of the
proposed control protocols are selected a priori, in accordance to the desired
transient and steady state performance specifications. In this way, the
selection of the control gains $\gamma_{i}, i\in\mathcal{V}$, that has been isolated
from the actual control performance, is significantly simplified to adopting
those values that lead to reasonable control effort. Nonetheless, it should be
noted that their selection affects both the quality of evolution of the
errors inside the corresponding performance envelopes as well as the control input
characteristics. 
Hence, fine tuning might be needed in real-time scenarios, to retain
the required control input signals within the feasible range that can be
implemented by real actuators. Similarly, the control input constraints impose
an upper bound on the required speed of convergence of $\rho^p_k(t)$, and $\rho^q_k(t), k\in\mathcal{M}$, as obtained by the exponentials $e^
{-l^p_kt},e^{-l^q_kt}$. Therefore, the selection of the control gains $\gamma_{i}$ 
can have positive influence on the overall closed loop system response. More specifically, notice that \eqref{eq:bar B_1}-\eqref{eq:e_bar_v} provide 
bounds on $\varepsilon,\varepsilon^v$ and $r,r^v$ that depend on the constants 
$\bar{B}_1,\bar{B}_2$. Therefore, in the special case that bounds on the model nonlinearities/disturbances are known, we can design the control gains $\gamma_i$ via \eqref{eq:u_i} 
such that the control signals $u_i$ are retained within certain
bounds.
\end{remark}

\begin{remark}
Regarding Assumption 1, we stress that, by  choosing the initial conditions $\theta_i(0),\forall i\in\mathcal{V}$ as well as the desired formation constants $\theta_{k,\text{des}} = q_{k_2,\text{des}},\forall k\in\mathcal{M}$ close to zero, the condition $-\tfrac{\pi}{2} < \theta_i(t) < \tfrac{\pi}{2}$ will not be violated, since the agents will be mostly operating near the point $\theta_i = 0, \forall i\in\mathcal{V}$. This is a reasonable assumption for real applications, since the angle $\theta_i$ represents the pitch angle of agent $i$ and is desired to be as close to zero as possible (consider, e.g., aerial vehicles). 

Furthermore, notice that the proposed control scheme guarantees collision avoidance only for the initially neighboring agents (at $t=0$), since that's how the edge set $\mathcal{E}$ is defined. Inter-agent collision avoidance with all possible agent pairs is left as future work by employing time-varying graphs. 
\end{remark}

\section{Simulation Results} \label{sec:simulation_results}
\begin{figure}[h!]
	\centering
	\includegraphics[scale = 0.65]{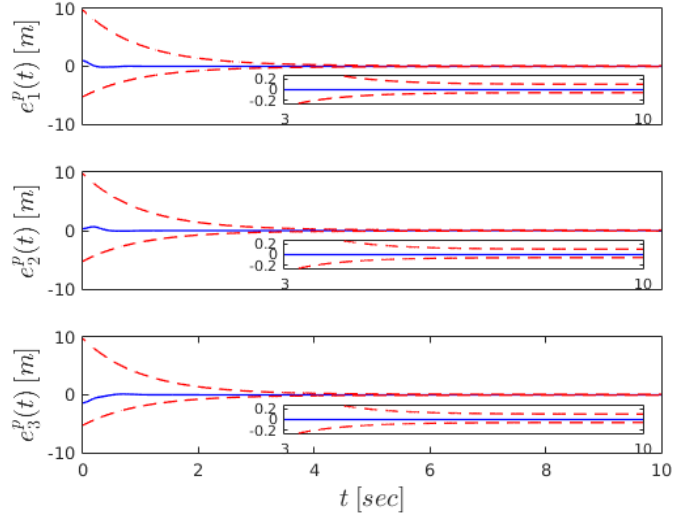}
	\caption{The evolution of the distance errors $e^p_k(t)$, along with the performance bounds imposed by $\rho^p_k(t), \forall k\in\{1,2,3\}$.}
	\label{fig:e_p}       
\end{figure}

\begin{figure}[h!]
\centering
  \includegraphics[scale = 0.65]{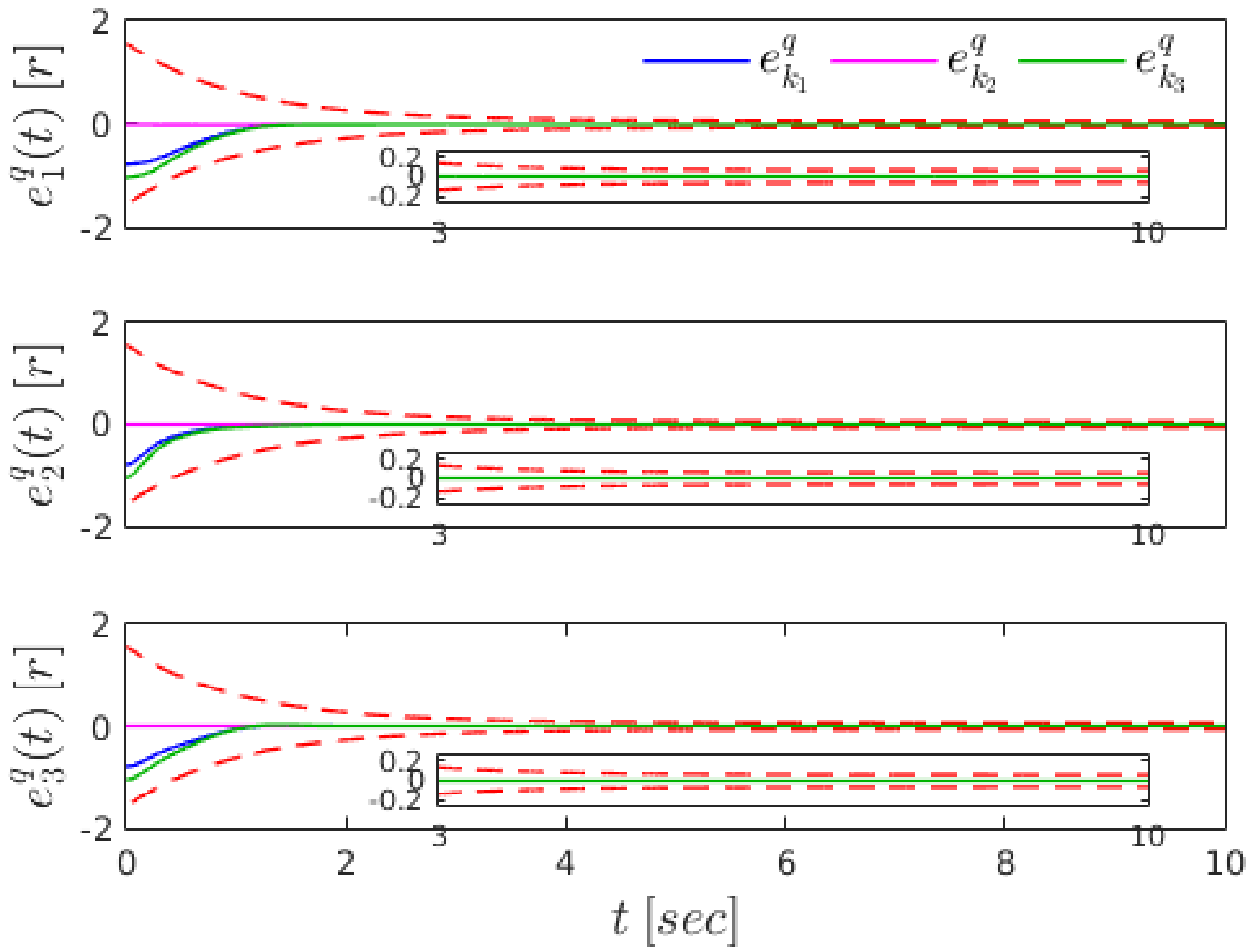}
\caption{The evolution of the orientation errors $e^q_{k_n}(t)$, along with the performance bounds imposed by $\rho^q_k(t), \forall k,n\in\{1,2,3\}$.}
\label{fig:e_q}       
\end{figure}

\begin{figure}
	\centering
	\includegraphics[scale = 0.65]{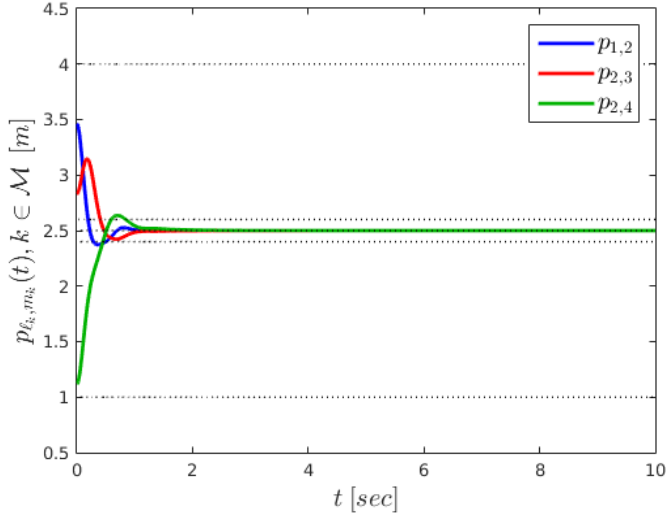}
	\caption{The distance between neighboring agents along with the collision and connectivity constraints.} 
	\label{fig:p}       
\end{figure}

\begin{figure}
\centering
  \includegraphics[scale = 0.65]{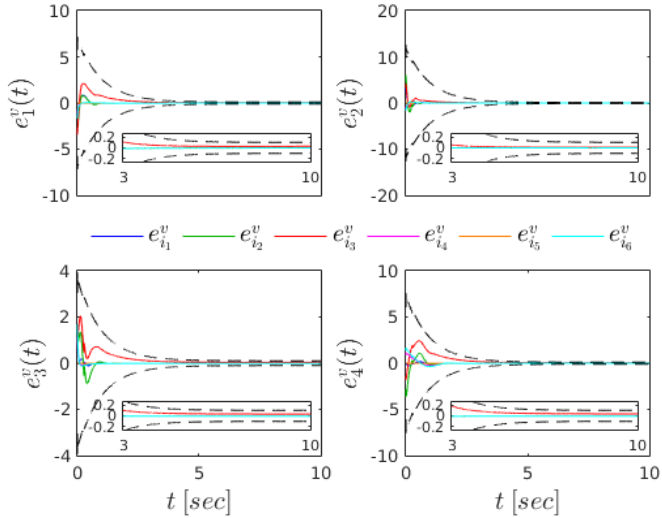}
\caption{The evolution of the velocity errors $e^v_{i_m}(t)$, along with the performance bounds imposed by $\rho^v_{i_m}(t), \forall i\in\{1,\dots,4\},m\in\{1,\dots,6\}$.}
\label{fig:e_v}       
\end{figure}

\begin{figure}
\centering
  \includegraphics[scale = 0.65]{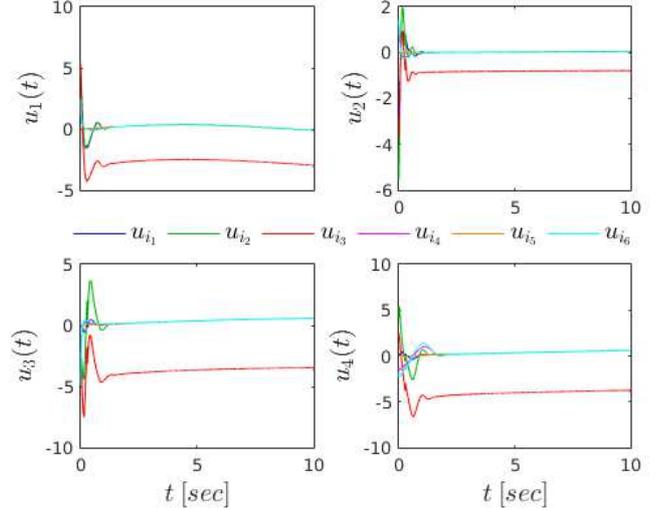}
\caption{The resulting control input signals $u_i(t), i \in \{1,\dots,4\}$.}
\label{fig:u}       
\end{figure}

To demonstrate the efficiency of the proposed control protocol, we considered a simulation example with $N=4, \mathcal{V} = \{1,2,3,4\}$ spherical agents of the form \eqref{eq:system}, with $r_i=1 \text{m}$ and $s_i = 4 \text{m},\forall i\in\{1,\dots,4\}$. We selected the exogenous disturbances as $w_i = A_i\sin(\omega_{c,i}t)(a_{i_1}x_i - a_{i,2}\dot{x}_i)$, where the parameters $A_i, \omega_{c,i}, a_{i_1}, a_{i_2}$ as well as the dynamic parameters of the agents were randomly chosen in $[0,1]$. The initial conditions were taken as $p_1(0)=[0,0,0]^T \ \text{m}, p_2(0)=[2,2,2]^T \ \text{m}, p_3(0)=[2,4,4]^T \ \text{m}, p_4(0)=[2,3,2.5]^T  \ \text{m}, q_1(0)=q_2(0)=q_3(0)=q_4(0) = [0,0,0]^T \ \text{r}$, which imply the initial edge set
 $\mathcal{E} = \{\{1,2\}, \{2,3\}, \{2,4\} \}$. The desired graph formation was defined by the constants $d_{k,\text{des}} = 2.5\text{m}, q_{k,\text{des}} = [\tfrac{\pi}{4},0,\tfrac{\pi}{3}]^T \ \text{r}, \forall k\in\{1,2,3\}$. Invoking \eqref{eq:C_k}, we also chose $C_{k,\text{col}} = 5.25\text{m}$ and $C_{k,\text{con}} = 10.75\text{m}$. Moreover, the parameters of the performance functions were chosen as $\rho^p_{k,\infty} = 0.1, \rho^q_{k,0} = \tfrac{\pi}{2} > \max\{e^q_{k_1}(0),e^q_{k_2}(0),e^q_{k_3}(0)\} = \tfrac{\pi}{3}$ and $l^p_k = l^q_k = 1, \forall k\in\{1,2,3\}$. In addition, we chose $\rho^v_{i_m,0} = 2\vert e^v_{i_m}(0) \rvert + 0.5, l^v_{i_m} = 1$ and $\rho^v_{i_m,\infty} = 0.1$. Finally, $\gamma_i$ is set to $5$ in order to produce reasonable control signals that can be implemented by real actuators. The simulation results are depicted in Fig. \ref{fig:e_p}-\ref{fig:u}. In particular, Fig. \ref{fig:e_p} and \ref{fig:e_q} show the evolution of $e^p_k(t)$ and $e^q_{k_n}(t)$ along with $\rho^p_k(t)$ and $\rho^q_k(t)$, respectively, $\forall k\in\{1,2,3\},n\in\{1,2,3\}$. Furthermore, the distances $\lVert p_{1,2}\rVert,\lVert p_{2,3}\rVert,\lVert p_{2,4}\rVert$ along with the collision and connectivity constraints are depicted in Fig. \ref{fig:p}. Finally, the velocity errors $e^v_{i_m}(t)$ along $\rho^v_{i_m}(t)$ and the control signals $u_i$ are illustrated in Figs. \ref{fig:e_v} and \ref{fig:u}, respectively. As it was predicted by the theoretical analysis, the formation control problem with prescribed transient and steady state  performance is solved with bounded closed loop signals, despite the unknown agent dynamics and the presence of external disturbances.
 \vspace{3mm}
\section{Conclusions and Future Work} \label{sec:conclusions}

In this work we proposed a robust decentralized control protocol for distance- and orientation-based formation control, collision avoidance and connectivity maintenance of multiple rigid bodies with unknown dynamic models. Simulation examples have verified the efficiency of the proposed approach. Future efforts will be devoted towards extending the current results to directed as well as time-varying communication graph topologies.


\bibliographystyle{plainnat}
\bibliography{ifacconf}             
                                                   
\appendix \label{app:proof_lemma}
\section{ }

\begin{lemma} \label{lem: P > 0}
	The matrix $P(x)$ is positive definite $\forall t\in[0,\tau_{\max})$.
\end{lemma}
\begin{pf}
	Firstly, note that Assumption \ref{assump:basic_assumption} implies that $\mathcal{G}$ is connected at $t=0$. Hence, in view of \eqref{eq:ksi_p_bounded}, $\mathcal{G}$ will stay connected for all $t\in[0,\tau_{\max})$. Moreover,
	since we do not consider adding edges to the graph,  $\mathcal{G}$ will also be a tree
	for all $t\in[0,\tau_{\max})$, and thus, the matrix $D^\tau(\mathcal{G})D(\mathcal{G})$ is positive definite for all $t\in[0,\tau_{\max})$, according to Lemma \ref{lemma:tree}. Therefore, the matrix 
	\begin{equation}
	\bar{D}^\tau(\mathcal{G})\bar{D}(\mathcal{G}) = \begin{bmatrix} 
	D^\tau(\mathcal{G})D(\mathcal{G})\otimes I_3 & 0_{3M\times 3M} \\
	0_{3M\times 3M} & D^\tau(\mathcal{G})D(\mathcal{G}) \otimes I_3
	\end{bmatrix}, \notag
	\end{equation}
	is also positive definite. Moreover, \eqref{eq:ksi_p_bounded} implies that $\lVert p_{\ell_k}(t) - p_{\ell_m}(t) \rVert > d_{k,\text{col}}, \forall t\in[0,\tau_{\max})$. Hence,
	there exists at least one $w\in\{x,y,z\}$ such that $(p_{\ell_k})_w(t) \neq (p_{\ell_m}(t))_w, \forall t\in[0,\tau_{\max})$, where $p_{\ell_a} = [(p_{\ell_a})_x,(p_{\ell_a})_y, \\ (p_{\ell_a})_z]^\tau, a\in\{k,m\}$. Therefore, $\text{rank}(\mathbb{F}_p(x)) = M$ and $\text{rank}(\bar{\mathbb{F}}_p(x)) = 4M$, which implies the positive definiteness of $P = \bar{\mathbb{F}}_p(x)\bar{D}^\tau(\mathcal{G})\bar{D}(\mathcal{G})\bar{\mathbb{F}}^\tau_p(x)$ (see Observation 7.1.8, pp. 431 in \citep{horn_jonshon}).
\end{pf}

\end{document}